\documentclass[prd,twocolumn,floatfix,preprintnumbers,reprint,showpacs]{revtex4}
\usepackage{epsfig,amsmath,amssymb,bm,color,graphicx}
\usepackage{afterpage}
\usepackage{pdfpages}

\usepackage{newtxtext,newtxmath}

\usepackage[caption=false]{subfig} 
\usepackage[T1]{fontenc}
\usepackage[normalem]{ulem}
\usepackage{lastpage}
\usepackage{soul}

\def\lya{Lyman-$\alpha$\,}
\def\mWDM{m_{\rm WDM}}

\def\fWDM{f_{\rm WDM}}

\newcommand{\be}{\begin{equation}}
\newcommand{\ee}{\end{equation}}

\usepackage{xcolor}

\usepackage[hidelinks]{hyperref}
\def\orcid#1{\href{https://orcid.org/#1}
{\includegraphics[keepaspectratio,width=0.7em]{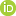}}}

\newcommand{\orcidauthorA}{\orcid{0000-0002-5445-461X}} 
\newcommand{\orcidauthorB}{\orcid{0000-0002-2642-5707}} 
\newcommand{\orcidauthorC}{\orcid{0000-0001-8443-2393}} 
\newcommand{\orcidauthorD}{\orcid{0000-0003-2764-8248}}

\begin{document}

\title{Constraining Mixed Dark Matter models with high redshift Lyman-alpha forest data}
\author{Olga Garcia-Gallego$^{1,2}$,\thanks{Email: og313@cam.ac.uk (OGG)}
Vid Ir\v{s}i\v{c}$^{1,3,4,5,6,7}$\,\orcidauthorA,\thanks{E-mail: v.irsic@herts.ac.uk (VI)}
Martin G. Haehnelt$^{1,2}$\,\orcidauthorC, 
Matteo Viel$^{4,5,6,7,8}$\,\orcidauthorB,
James S. Bolton$^{9}$\,\orcidauthorD,
}

\smallskip
\affiliation{
$^{1}$Kavli Institute for Cosmology, University of Cambridge, Madingley Road, Cambridge CB3 0HA, UK\\
$^{2}$Institute of Astronomy, University of Cambridge, Madingley Road, Cambridge CB3 0HA, UK\\
$^{3}$Center for Astrophysics Research, Department of Physics, Astronomy and Mathematics, University of Hertfordshire, College Lane, Hatfield AL10 9AB, UK\\
$^{4}$SISSA- International School for Advanced Studies, Via Bonomea 265, 34136 Trieste, Italy\\
$^{5}$INFN – National Institute for Nuclear Physics, Via Valerio 2, I-34127 Trieste, Italy\\
$^{6}$IFPU, Institute for Fundamental Physics of the Universe, via Beirut 2, 34151 Trieste, Italy\\
$^{7}$INAF, Osservatorio Astronomico di Trieste, Via G. B. Tiepolo 11, I-34131 Trieste, Italy\\
$^{8}$ICSC - Centro Nazionale di Ricerca in High Performance Computing, Big Data e Quantum Computing, Via Magnanelli 2, Bologna, Italy\\
$^{9}$School of Physics and Astronomy, University of Nottingham, University Park, Nottingham, NG7 2RD, UK\\
}

\begin{abstract}
This study sets new constraints on Cold+Warm Dark Matter (CWDM) models by leveraging the small-scale suppression of structure formation imprinted in the Lyman-$\alpha$ forest. Using the Sherwood-Relics suite, we extract high-fidelity flux power spectra from simulated Lyman-$\alpha$ forest data, spanning a broad range of cosmologies and thermal histories. This enables precise constraints on the warm dark matter (WDM) fraction, $\fWDM$, and the mass of the WDM particle, $\mWDM$. A key advancement of our analysis is the integration of a neural network emulator directly at the likelihood level, significantly accelerating Bayesian parameter inference. With new observations of high-redshift ($z$ = 4.2$-$5.0) quasar spectra from UVES and HIRES, we establish stringent upper limits: for $\mWDM$ = 1 keV, we find $\fWDM < 0.16$ (2$\sigma$), with constraints loosening to 35\%, 50\%, and 67\% for $\mWDM$ = 2, 3, and 4 keV, respectively. Our results for pure WDM reaffirm the lower bounds of previous work. Crucially, we account for the fixed resolution of simulations and the impact of patchy reionization, demonstrating their minimal influence on mixed dark matter constraints. This robustness paves the way for tighter bounds with improved statistical samples in the future. Our findings suggest that CWDM models can naturally accommodate mild suppression of matter clustering in the high redshift Lyman-$\alpha$ forest 1D flux power, potentially offering a resolution to some of the ongoing cosmological tensions at low redshifts, namely the $S_{8}$ tension.
\end{abstract}

\maketitle
\section{Introduction}

In the standard cosmological model, $\Lambda$CDM, cold non-baryonic dark matter with negligible primordial velocities constitutes approximately $\Omega_{\mathrm{CDM}}$ $\approx$ 0.26 of the total energy density of the Universe. This model provides a pivotal framework for galaxy formation theory, where the hierarchical merging of non-linear dark matter structures, or haloes, plays a key role in shaping the large-scale structure of the Universe. In this context, the $\Lambda$CDM paradigm accurately predicts the evolution of the matter power spectrum across cosmic time from CMB observations \cite{Planck2021}, demonstrating remarkable success in matching observations down to scales comparable to the average spacing between galaxies ($\approx$ a few Mpc). 

At smaller scales, however, discrepancies arise between theoretical predictions from N-body simulations and observational data. Some of these inconsistencies stem from the hierarchical nature of the theory. While the apparent mismatch in the number of dwarf galaxies in the Milky Way (\textit{Missing Satellites Problem}, \cite{Klypin1999, Moore1999}), has been largely resolved through new observations of faint satellites (\cite{Simon19}) and improved baryonic modelling in high-resolution simulations (e.g. \cite{Engler21, Font21}), other tensions persist. The most significant among these is related to the amount of dark matter in the inner parts of galaxies (\textit{Cusp-Core Problem},  \cite{Blok2010}), particularly the challenge of explaining the diverse rotation curves observed \cite{Santos20}. Moreover, while the dynamics of the most massive Milky Way satellites (\textit{Too-Big-To-Fail Problem}, \cite{Boylan-Kolchin2011, Boylan-Kolchin2012}) are now better understood through updated cosmological simulations (e.g. \cite{Wetzel16, Samuel20}), the situation remains unclear for satellites outside the vicinity of the Milky Way (\cite{Sales22}). 

Another key challenge of CDM (cold dark matter) on small scales involves the amplitude of matter density fluctuations, quantified by $\sigma_{8}$. The parameter combination $ S_{8} = \sigma_{8} \sqrt{\Omega_{\text{m}}/0.3}$ has given rise to the so-called $S_{8}$ tension, where cosmic shear, galaxy clustering surveys, and their cross-correlation with CMB lensing (e.g. \cite{White22, Ferraro22, Amon23, Ivanov23, Lange23, Farren24, SFChen24})  suggest lower values than those inferred from the CMB (\cite{Planck2021}).

A major open question in cosmology is whether both of these problems could be solved by baryonic physics (e.g. \cite{Mashchenko06, Brooks13, Fattahi16, Pontzen14, Daalen11, Volker18, Chisari19}) or dark matter models beyond the Standard Model (e.g. \cite{Su11, Wang17, Lovell17, Liu18, Nadler23, Lague2024, Rogers25}). 

Proper modeling of the former, however, is computationally challenging, and while the physics of galaxy formation is understood in broad terms, many of the underlying baryonic processes, such as feedback mechanisms, remain poorly constrained \cite{vanDaalen11, Thorsten17, Vogelsberger20, Medlock25, Bigwood2025}. As such, exploring alternative dark matter models is further motivated by two primary considerations: (1) the absence of a viable dark matter candidate within the Standard Model of particle physics; (2) no direct detection has been achieved thus far, after a decade of experiments aimed at detecting either of the two most popular CDM candidates: WIMPs (weakly interacting massive particles) (LHC), and axions (ADMX, \cite{ADMX}).

From a particle physics perspective, the simplest extension of the Standard Model is to accommodate a \textit{heavy} or \textit{sterile} neutrino that can explain the mechanism facilitating neutrino oscillations \cite{Boyarsky19}. Sterile neutrinos can be produced via active neutrinos oscillations in the early Universe, as proposed by \cite{Dodelson1994}. They would be the prototype for  Warm Dark Matter (WDM), an hypothetical thermal relic dark matter particle that decouples earlier and is more massive than Standard Model neutrinos, initially proposed in \cite{Blumenthal1982, Peebles1982, Bond1982}. Such a dark matter candidate has a non-negligible primordial velocity distribution at the time of decoupling, and can consequently escape small gravitational perturbations, suppressing structure formation on scales smaller than their free-streaming length, $\lambda_{\rm fs}$, which is inversely proportional to the WDM particle mass. For WDM particle masses of the order of a few keV, the free-streaming length roughly corresponds to the size of a dwarf galaxy. WDM follows the same bottom-up structure formation as CDM, making it a promising candidate to address small-scale challenges while preserving the successes of CDM at larger scales. The most recent constraint from cosmological analysis on this model excludes thermal relics lighter than $5.7\,$ keV (2$\sigma$) \cite{Irsic2023}.

For this reason, the interest in a hybrid model that interpolates between CDM-like and WDM cosmology has re-emerged. The Cold$+$Warm Dark Matter (CWDM) model was first introduced by \cite{Davis1992, Taylor1992, Dalen1992, Klypin1993}. The model is parameterized by the fraction of the warm (\textit{lighter}) counterpart, $f_{\text{WDM}} = \Omega_{\text{WDM}}/ \Omega_{\text{M}}$; and its particle mass $\mWDM$, which has the same properties as in a pure WDM cosmology. 

In some variants of the production mechanism of sterile neutrinos, the resulting velocity dispersion of the hypothetical particle consists of cold and warm components \cite{Shi1999, Boyarsky2009}. Within this framework, a structured dark matter sector could exist, featuring components in the keV and GeV mass regimes, making sterile neutrinos viable candidates for CWDM \cite{Shaposhnikov2006}.
Such a dual-component nature of dark matter has also been proposed for axions \cite{Lague2024}, reinforcing the motivation to investigate CWDM.

The free-streaming property of CWDM can be constrained observationally from the Lyman-$\alpha$ forest,  a powerful cosmological probe of matter density fluctuations in the weakly non-linear regime through the high redshift and underdense IGM. In the past, the \lya forest has been used, combined with CMB data, to constrain the matter power spectrum (e.g. \cite{Croft2002, Pat2005, Lidz2010}), nature of dark matter \cite{Viel13wdm,Irsic2017b,Irsic2017c,Palanque-Delabrouille2020,Garzilli2021,Rogers2021}, and to test physics beyond the Standard Model using the 3D correlation function to measure the scale of Baryon Acoustic Oscillations (e.g. \cite{Slosar2011, Slosar2013, Busca2013, Bourboux2020, DESIDR1, DESIDR2}).

At the scales of interest for the dark matter models, the \lya forest is influenced by two key factors: the thermal evolution of the intergalactic medium (IGM) (e.g., \cite{Boera2019}) and the underlying dark matter density field, which drives structure formation through gravitational interactions (e.g., \cite{Villasenor2023, Irsic2023}). Consequently, accurately interpreting observations of the \lya forest at these scales requires hydrodynamic simulations to model the complex gas dynamics. These simulations have facilitated detailed comparisons between the structures predicted by the CDM paradigm and those observed in the \lya forest down to the Jeans scale (\cite{Viel2009}), while also incorporating the free-streaming properties of dark matter.
In the past, constraints on the mass $\mWDM$ have been obtained by comparing simulations with quasars (QSOs) datasets of different resolution and signal-to-noise ratios (S/N). This comparison is usually done within a Bayesian framework by interpolating in the parameter space where simulations are available (\cite{Viel2013a, Irsic2017b, Yeche2017, Palanque-Delabrouille2020, Molaro2021}) , which become unfeasible as the dimensionality of the inference problem increases. Instead, surrogate models or emulators can be used based on Gaussian process interpolation (\cite{Bird2019, Walther2019, Rogers2019, Walther2021, Fernandez2022, Bird2023}), or a supervised neural network interpolator (\cite{Cabayol-Garcia2023, Molaro23}), demonstrating significant inference speedup. 
The data used to carry out this comparison has included, in increasing order of resolution, the Sloan Digital Sky Survey (SDSS) and the Baryon Oscillation Spectroscopic Survey (BOSS) (e.g., \cite{Baur2017, Palanque-Delabrouille2020}), X-Shooter (e.g., \cite{Irsic2017b}) and MIKE spectrograph at the Magellan Telescopes (e.g. \cite{Viel2013a}), and HIRES spectrograph at the Keck I Telescope (Keck) and UVES spectrograph of the Very Large Telescope (VLT) (e.g. \cite{Irsic2023}). The higher-resolution data provides stronger bounds on $\mWDM$ by probing the small scales where the suppression occurs.

Beyond pure WDM, \cite{Palazzo2007} first suggested a preference for $\fWDM$ $<$ 0.7 mixed models by reinterpreting Lyman-$\alpha$ forest and X-ray data in the presence of sterile neutrinos. Later, \cite{Boyarsky2009} combined the WMAP5 and Lyman-$\alpha$ (SDSS) data to constrain $\fWDM$ $<$ 0.35 for the minimum mass of their prior range $\mWDM \sim$ 1 keV in the case of thermally decoupled candidates. \cite{Baur2017} found that a mass as light as $\mWDM$ $>$ 0.7 keV is allowed if $\fWDM$ $<$ 0.15. \cite{Diamanti17} combined Planck, baryon acoustic oscillations and Milky Way satellites data to constrain $\fWDM$ $<$ 0.29 for masses in the range 1-10keV. \cite{Parimbelli22} used N-body similations and model the effect of baryons on top to predict allowed mixed models to which weak lensing and galaxy clustering power spectra upcoming data will be sensitive to. 

In this work, we update previous constraints on CWDM models using high-$z$ Lyman-$\alpha$ forest measurements that are sensitive to matter clustering to 2x as small separations as any previous attempts, $k \sim 20$ $h\,\mathrm{Mpc}^{-1}$ \cite{footnote}; and a set of hydrodynamical simulations that include the effect of a heavy and a light thermal dark matter particle in structure formation including full gas physics. We compare the synthetic spectra to the data in a Bayesian inference framework. Since the grid of simulations spans a high-dimensional parameter space, we efficiently sample the likelihood by implementing a neural network emulator. We perform a variety of analyses that incorporate numerical correction due to a fixed resolution of simulations (resolution correction), inhomogeneous reionization effects and a possible mis-estimation of instrumental noise of the data. 

The structure of this paper is as follows: In Section~\ref{section1}, we present the 1D Lyman-$\alpha$ flux power spectrum measurements. Section~\ref{section2} introduces the hydrodynamical simulations, highlighting their key properties and the impact of CWDM on the matter power spectrum. We also describe the synthetic flux power spectrum models derived from these simulations. Section~\ref{section3} details the emulator used for the MCMC analysis. In Section~\ref{section4}, we discuss  various analysis choices, comparing their effect on constraints on mixed dark matter models and exploring the degeneracy between the WDM parameters $\fWDM$ and $\mWDM$. Finally, Section~\ref{section5} summarizes our findings and examines the implications of the allowed CWDM models for the $S_{8}$ tension.

\section{Data}\label{section1}

We utilize the same dataset as presented in \cite{Irsic2023}, specifically the one-dimensional (1D) flux power spectra introduced in \cite{Boera2019}. These spectra were derived from a sample of 15 high-resolution quasar observations conducted with the UVES and HIRES spectrographs per averaged redshift bin, centered at $z$ = 4.2, 4.6 and 5.0. In particular, we analyze the power spectra across 16 $k$-bins, uniformly spaced by $\Delta_{\mathrm{log_{10}} (k)}$=0.1, from $\mathrm{log_{10}} (k/\mathrm{km^{-1}\,s})$ = -2.2 to $\mathrm{log_{10}} (k/\mathrm{km^{-1}\,s})$ = -0.7. The measurements at these $k$-bins incorporate corrections for finite spectral resolution and pixel size, as detailed in Appendix K of \cite{Boera2019}. These corrections were derived by applying the same window function as in \cite{Palanque-Delabrouille15, Irsic2017a, Boera2019}. Furthermore, the non-diagonal elements of the covariance matrix are adopted from \cite{Irsic2023}, which employed the same box size, $L_{\text{box}}$, as in this work to construct the bootstrap cross-correlation coefficient.

The velocity width corresponding to the smallest scale probed in this dataset is approximately 30 km$\,\mathrm{s}^{-1}$, which is significantly larger than the spectral resolutions of HIRES and UVES, measured as 6 $\approx$ km$\,\mathrm{s}^{-1}$ and 7 $\approx$ km$\,\mathrm{s}^{-1}$, respectively. These characteristics make it an ideal dataset to resolve the small-scale features of the Lyman-$\alpha$ forest. Prior studies have used lower-resolution datasets with a larger number of QSO samples (e.g., \cite{Yeche2017, Palanque-Delabrouille2020}) or data from the same instruments but with approximately half the number of quasar spectra (e.g., \cite{Viel13, Irsic2017a}). These studies have been limited to scales of up to $k \approx 0.1$ s$\,\mathrm{km}^{-1}$. While such datasets have been effective in constraining both the instantaneous thermal state and the evolution of the intergalactic medium (IGM), \cite{Boera2019} extends this reach to $k_\mathrm{max} = 0.2$ s$\,\mathrm{km}^{-1}$.  Crucially, the high $k$-bins introduced in this work are the ones sensitive to dark matter free-streaming, allowing us to probe power suppression on scales of 10–100 kpc. This extension to smaller scales is made possible by improved modeling of instrumental noise and metal contaminants \cite{Day19, Boera2019, Karacali22}. To account for these uncertainties, we marginalize over these effects as part of our analysis.

\section{Simulations}\label{section2}

\begin{table*}[htbp!]
\centering

\begin{tabular}{lcccccccc} 
\hline
Name     & $L_{\rm box}$ & $N_{\rm part}$ & $z_{\rm rei}^{\rm end}$ & $T_{0}(z=4.6)$ & $u_{0}(z=4.6)$ & WDM mass & $\fWDM$\\
 & $[h^{-1}\rm\,cMpc]$ & &  & [$\rm K$] &  $[\rm eV\,m_{\rm p}^{-1}]$ & $[\mathrm{keV}^{-1}]$   \\
\hline 

L20-ref      & 20.0 & $2\times 1024^{3}$ & 6.00 & 10,066 & 7.7 & $[0,\frac{1}{4},\frac{1}{3},\frac{1}{2}, 1]$ & $[0,\frac{1}{8},\frac{1}{4}, {\frac{1}{2}}, 1]$  \\
\hline 
L20-ref$^{\,\mathrm{CLASS}}$      & 20.0 & $2\times 1024^{3}$ & 6.00 & 10,066  & 7.7 & $0$ & $0$  \\
\hline 
R10-ref      & 10.0 & $2\times$[$1024^{3}$,$512^3$] & 6.00 & 10,066  & 7.7 & $[0,\frac{1}{4},\frac{1}{3},\frac{1}{2}]$ & [0, 1] \\

\hline
\end{tabular}
 \caption{Reference thermal history simulations used in this work. A complete list, including all thermal history runs, can be found in Table I of \cite{Irsic2023}.  The columns include the box size ($[h^{-1}\rm\,cMpc]$), the number of particles, the end of reionisation redshift, the thermal parameters $T_{0}$ and $u_{0}$ at redshift $z$ = 4.6, the inverse of the WDM mass ${\mWDM}^{-1}$, and the additional dark matter parameter, the fraction of WDM particle mass $\fWDM$. The first two rows list the main simulations: one with fiducial CAMB initial conditions (as in \cite{Irsic2023}, \cite{CAMB}) and the other using the CLASS code (\cite{CLASS}). The last row contains the fiducial simulation for mass resolution correction. Box correction runs match those in \cite{Irsic2023}.  Note that the simulations in all three rows extend to $\mWDM$ = 1 keV, with the first set covering CWDM models. Different CWDM models listed in this table were run for all 12 thermal histories of \cite{Irsic2023}.} \label{sims}
\end{table*}

To probe CDM and pure warm dark matter WDM cosmologies, we use the same set of simulations from the Sherwood project as employed in \cite{Irsic2023}, incorporating the fiducial ultraviolet (UV) background from \cite{Puchwein2019}. These simulations feature several advancements over previous models used in earlier studies (\cite{Viel2013a, Irsic2017b}), including the implementation of a non-equilibrium thermo-chemistry solver.

\begin{figure}[b]
  \hspace{0.1\textwidth}
 \centering
 \includegraphics[width=\linewidth]{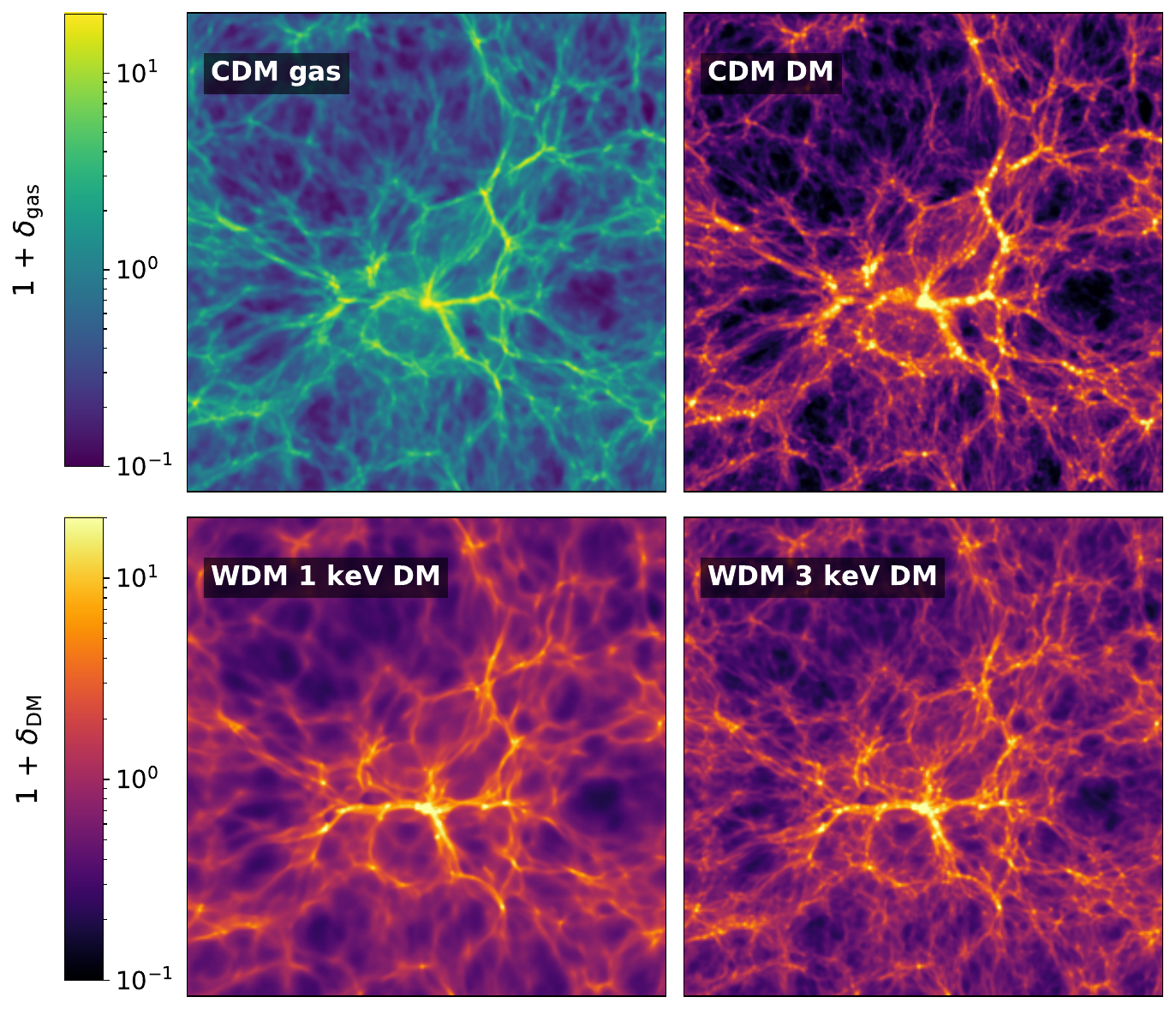} 
\caption{Slice through a cosmological simulation of size 20 $h^{-1}$\,Mpc with 2 $\times$ 1024$^3$ dark matter (DM) and gas particles at $z$ = 4.2 from  the Sherwood-Relics simulation suite that was run with the \texttt{P-Gadget3} code. Top panel: Hydrogen gas distribution (left), dark matter distribution in the standard CDM model (right). Bottom panel: dark matter distribution in a 1\,keV WDM model (left) compared to a 3\,keV WDM model (right) with brighter colours indicating higher density. } \label{slice}
\end{figure} 

\begin{figure*}[htbp!]
    \centering
    \begin{minipage}{0.43\linewidth}
        \centering
        \includegraphics[width=\linewidth]{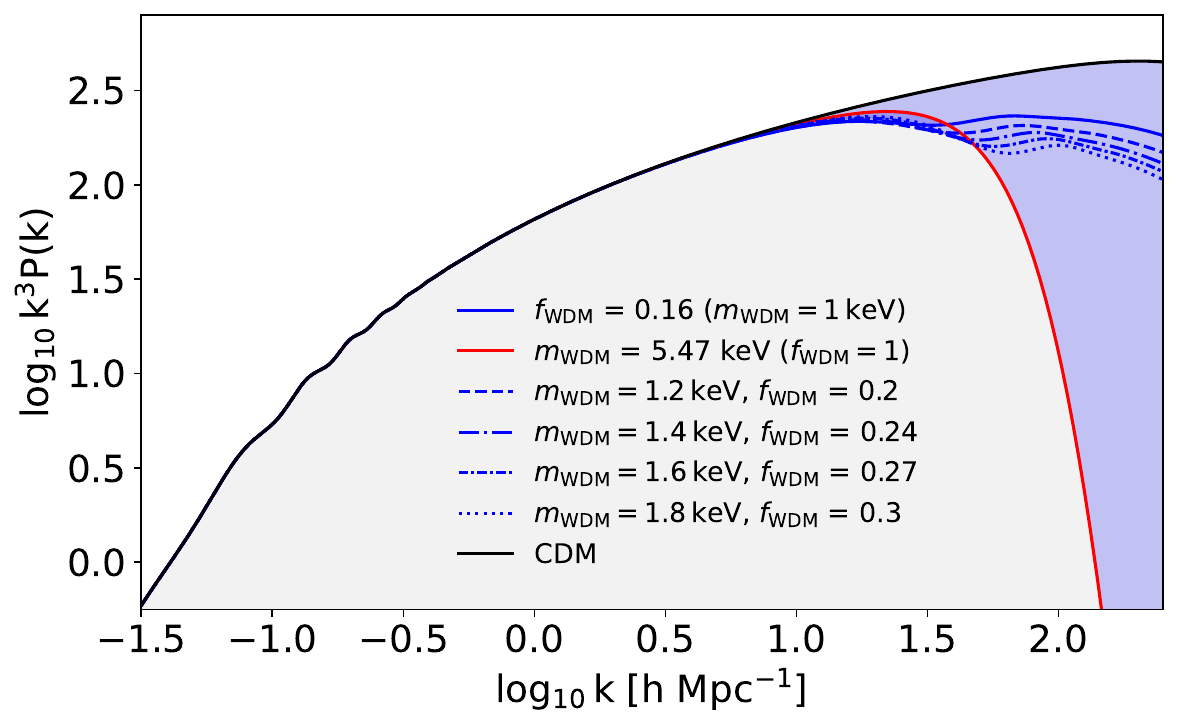}
    \end{minipage}
    \hfill
    \begin{minipage}{0.55\linewidth}
        \centering
        \includegraphics[width=\linewidth]{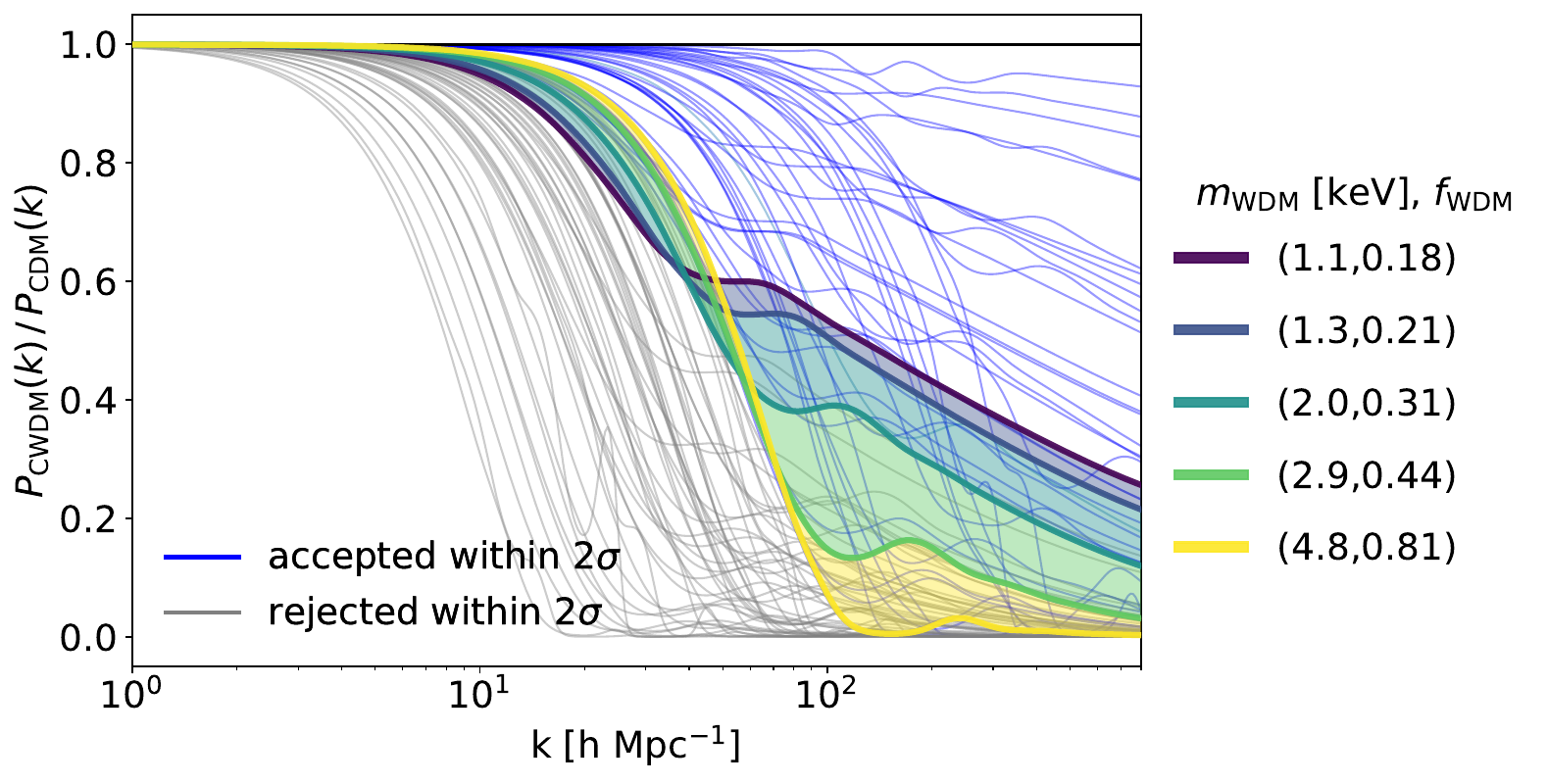}
    \end{minipage}
    \caption{\textit{Left}: Linear matter power spectrum at $z$ = 0 computed using the Boltzmann solver CLASS for CDM (black), a CWDM model (blue) and pure WDM model (red). The last two correspond to the 2$\sigma$ bounds found when fixing the highest $\fWDM$ ($\fWDM$ = 1) and lowest $\mWDM$ ($\mWDM$ = 1\,keV) in our analysis. We also present a range of CWDM models, shown in blue with varying line styles, which are drawn from the region of power suppression of allowed models. These models delineate the 2$\sigma$ allowed region, represented by the blue shaded area. The region of rejected models is shown in gray. \textit{Right}: The matter transfer function ($T^2(k)$) from random samples of the 2D posterior distribution in the (${\mWDM}^{-1}-\fWDM$) plane inside (outside) the 2$\sigma$ contour in blue (gray). We show five allowed models in color, highlighting that we are sensitive to both the CWDM departure scale from CDM (mainly set by $m_{\mathrm{WDM}}$) and the suppression amplitude (dependent on $f_{\mathrm{WDM}}$). Notably, lower $m_{\mathrm{WDM}}$ values can be constrained at the cost of reduced suppression at higher-$k$. }
    \label{matterPS}
\end{figure*}

In this work, we expand upon the Sherwood simulations framework by conducting additional simulation runs aimed at constraining CWDM models detailed in Table~\ref{sims}. The simulations are performed within a cosmological volume of 20 $h^{-1}\,$Mpc, with 2$\times$1024$^3$ dark matter and gas particles, which provides sufficient resolution to capture the small-scale structure of the \lya forest (\cite{Irsic2023}) (see Subsection~\ref{Rs}). Further details regarding the simulations can be found in \cite{Irsic2023} (Table I) and \cite{Puchwein2023}.

In  Figure~\ref{slice} we show the projected gas and dark matter density fields extracted from simulation runs for CDM and WDM cosmologies. One can see that the overdense regions in the gas distribution, which serve as the seeds for the formation of stars and galaxies, closely follow the distribution of dark matter on large scales. The gas distribution at the small scales appears more diffuse compared to the dark matter distribution in the top-right panel due to the pressure smoothing effect: the gas hydrodynamically reacts to heating from reionization by expanding, suppressing small-scale structure \cite{Puchwein2019}. The neutral hydrogen, which is responsible for the Lyman-$\alpha$ absorption, resides in these filaments or voids. The bottom two panels highlight the primary effect of dark matter free-streaming on structure formation. As $\mWDM$ decreases in these pure WDM models, the free-streaming length increases. This results in the smoother filamentary structure observed in the Figure. Notably, this smoothing effect is more pronounced for the $\mWDM = 1$ keV model compared to the $\mWDM = 3$ keV model.

\subsection{Imprint of mixed dark matter models on the matter power spectrum}

The initial conditions for WDM and CWDM simulations differ from those of CDM in the computation of the power spectrum. The free-streaming effect, $\lambda_{\rm fs}$, is incorporated through a transfer function, $T(k)$, that modifies the original power spectrum, (e.g. \cite{Viel2005}). Specifically, the power spectra for CDM and pure warm thermal relic models are computed using the Boltzmann solver CAMB \cite{CAMB}, while for the latter cosmology, we apply the transfer function from \cite{Bode2001, Viel2005, Viel2013}. The mixed models are generated using the in-built CLASS option, which accounts for the temperature, mass, and density of non-cold dark matter \cite{CLASS, Lesgourges2015}. To ensure consistency across different cosmologies simulated with different codes, we correct for minor discrepancies by taking the ratio of the flux power spectrum between CDM models computed with CAMB and CLASS (see Table~\ref{sims}).

In the context of CWDM models, the transfer function asymptotically approaches a constant plateau at small scales. The height of this plateau is primarily determined by the warm dark matter (WDM) fraction, $\fWDM$, due to the presence of a dominant CDM component. The characteristic cut-off scale is governed by the mass of the warm dark matter counterpart, similar to the behavior observed in pure WDM models.

Figure~\ref{matterPS} shows on the left the linear matter power spectrum for CDM at $z=0$, along with the 2$\sigma$ lower bound on pure WDM and on a mixed model with $\mWDM$ = 1 keV. The latter is the lightest thermal relic allowed in our analysis. We further show models with progressively heavier warm counterparts (and correspondingly smaller fractions), which bracket the 2$\sigma$ allowed region found in this work, shaded in blue. This Figure illustrates the imprint of CWDM in the matter power spectrum: the scale at which it deviates from CDM is set by $\mWDM$ while the suppression depends on the abundance given by $\fWDM$. We show in the right plot of Figure~\ref{matterPS} the distinct suppression patterns of mixed CWDM models relative to CDM. In particular, these hybrid models can remain light as long as $\fWDM$ is small, resulting in a weaker suppression of the matter power spectrum at slightly larger scales. 

However, CWDM models with stronger suppression at larger scales also show less suppression at small scales, compared to pure WDM models. This is shown for a collection of hand-picked models in the color gradient of Figure~\ref{matterPS}. For such models we find that the Lyman-$\alpha$ forest data in this work constrain both the scale of the small-scale suppression, and the shape of the transfer function around the suppression scale.

\subsection{Flux power spectrum models} \label{flux_models}
\begin{figure*} [htbp!]
\centering
\includegraphics[width=0.7\linewidth]{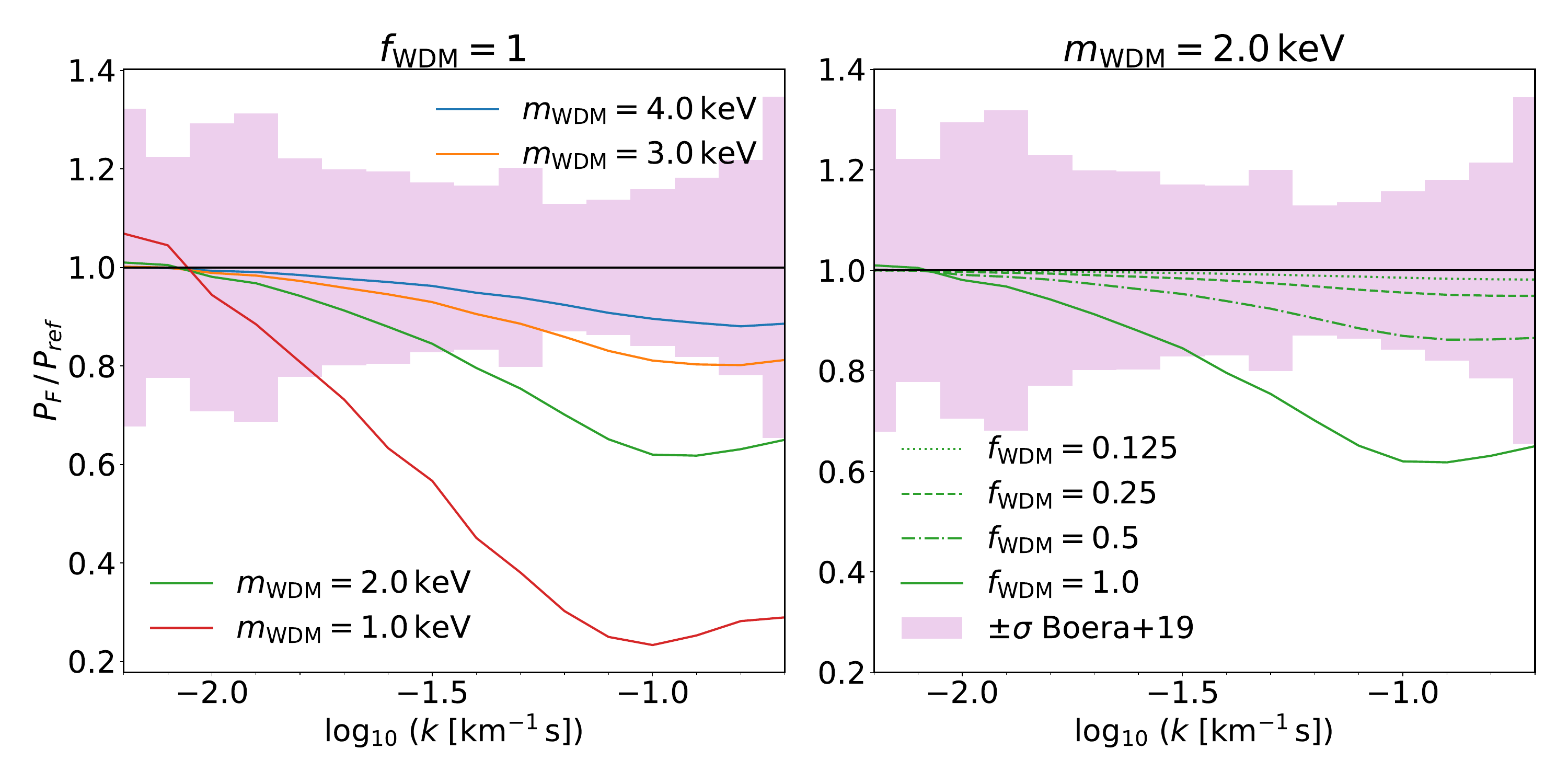}
\caption{Flux power spectrum ratio of simulated models to the reference model for CDM cosmology with post-processed parameters fixed to fiducial values ($T_{0} = 10700\,$K, $\gamma$ =  1.2) at $z$ = 4.2. \textit{Left}: The WDM particle mass  varied with fixed $f_{\text{WDM}}$ = 1. \textit{Right}: CWDM models with vaying $f_{\text{WDM}}$ for $m_{\text{WDM}} = 2\,$keV. } 
\label{fig:models}
\end{figure*}

The synthetic Lyman-$\alpha$ forest spectra, specifically the optical depth field ($\tau$), can be computed by extracting the information on neutral hydrogen number density, temperature and peculiar velocity contained in each pixel in the box along the different lines-of-sight. These are parallel to the boundaries of the box and have  $N_{\text{bins}}$ = 2048 number of bins with width $\delta x$ = $L_{\text{box}}/N_{\text{bins}}$. We extract $N_{\text{LOS}}$ = 5000 lines-of-sight and compute the flux  ($F$ = $e^{-\tau}$) field for each. The perturbations in flux, $\delta_{\text{F}}=F/{\bar F}-1$, are characterized by the 1D flux power spectrum,  $P_{\text{F}}(k)$ = $L^{-1}_{\text{box}}{\langle{\delta_{\text{F}}}{\delta^{*}_{\text{F}}}\rangle}$, where $\bar{F}$ is the mean transmitted flux calculated as an average of the flux field across all the pixels extracted from a simulation at each redshift. We  repeat this for the three redshift bins used in our analysis.

Without post-processing, the number of models that we have established are spanned by the allowed WDM parameters, $m_{\text{WDM}}$ = [1,2,3,4] keV and $\fWDM$ = [0.125,0.25,0.5,1.0]. These add up to 16 WDM cosmologies + 1 from CDM ($\fWDM$ = 0). As described above, for each cosmology, we run 12 simulations, each one with different thermal history, characterized by the evolution of the cumulative injected heat into the IGM at mean density ($u_0(z)$). This amounts to 17$\times$12 = 204 models per redshift bin.

Under the assumption of photo-ionization equilibrium, we post-process the simulations before inferring our cosmology from the flux power spectra models in two steps. First, we exploit the temperature dependence of the recombination coefficient $\alpha_{\text{HII}}$ $\propto$ T$^{-0.72}$ in the Lyman-$\alpha$ forest regime to isolate the impact of the instantaneous temperature of the gas on the power spectrum. We do this by rotating and translating the $T-\rho$ plane of simulation parameters as has been done in previous work (e.g. \cite{Boera2019}, \cite{Irsic2023}), recalculating the $\tau$ field for a range of $T_{0}$, $\gamma$ values. In this way, we obtain flux models for a combination of $T_{0} \in$ [5000, 14000]$\,$K spaced by 1000$\,$K, and $\gamma \in$ [0.9, 1.8] in steps of 0.1. 

In the second step, we adopt the common approach of adjusting the mean flux, $\bar{F}$, derived from the simulated spectra to match the observed measurements of effective optical depth $\tau_{\text{eff}}$, in order to account for uncertainties in the UV background. This is done by rescaling the $\tau$ field , with $\bar{F}$ defined as $\bar{F} = e^{-\tau_{\text{eff}}}$, since $\tau \propto \Gamma_{\text{HI}}^{-1}$ under the photo-ionization equilibrium assumption. This method is preferred over direct rescaling of $\Gamma_{\text{HI}}$ because the mean transmitted Lyman-$\alpha$ flux is measurable (\cite{Becker13, Bosman18, Bosman22}), providing valuable prior knowledge of $\tau_{\text{eff}}$ to calibrate simulations and infer physically motivated posteriors on model parameters.

 
After post-processing, the total number of models span a grid of 15 $\times$ 10 $\times$ 10 $\times$ 204 models = 306000 parameter values.

In Figure~\ref{fig:models}, we show the relative ratio of these simulated flux power spectrum for WDM and CWDM cosmologies to the reference CDM simulation at $z$ = 4.2 and for a fixed thermal history ($u_{0} = 8.14$\,eV, $T_{0} = 10700\,$K, $\gamma$ =  1.2) and the mean flux (${\bar F}$) is fixed such that the effective optical depth matches that of \cite{Boera2019}, i.e. $\tau_{\mathrm{eff}}$ $\equiv -\ln{{\bar F}}$ $=$ ${\tau_{\mathrm{eff}}}^{\mathrm{B}}$. 
One can see that the suppression in power at small scales (high $k$ values) is enhanced for WDM models with lower $m_{\text{WDM}}$. This agrees with what one  expects for a lighter DM particle that has a larger free streaming length. The magnitude of the suppression increases as the larger free streaming strength  suppresses structure at even larger scales. This can also be seen in the behaviour of the power spectrum for different $\fWDM$ values, namely, the smaller $\fWDM$, the weaker the effect on the power spectrum, eventually recovering CDM. The 1$\sigma$ uncertainties from \cite{Boera2019} are also shown to illustrate the sensitivity of the data to the WDM models. The sensitivity is higher for "hotter" dark matter models, whereas the "colder" dark matter models differ by less than the error margin, making it difficult for the data to distinguish between them effectively.

\subsection{Mass resolution and box size correction} \label{Rs}
Prior to performing the inference, we test whether the models extracted from simulations are able to resolve the Lyman-$\alpha$ forest structure by running numerical convergence tests. The two relevant parameters are $L_{\text{box}}$ and $N_{\text{p}}$ (\cite{Borde2014, Villasenor2021, Doughty2023}). 
The first parameter sets the number of $k$-modes on large scales and influences all scales due to non-linear coupling between the modes. When $L_{\text{box}}$ is kept fixed, the smallest resolvable scale is governed by the particle number, $N_{\text{p}}$.

Following \cite{Irsic2023}, we splice together the flux power from large and small box size simulations, $L_{\text{box}}$ = 40 and 10\,$h^{-1}\,$Mpc, varying $\,N_{\text{p}}$ = [1024$^{3}$, 512$^{3}$] (\cite{McDonald03}). We check that \cite{Boera2019} is sensitive only to the intermediate splicing regime. Therefore, we correct the flux power spectra obtained with our reference simulation $P_{20, 1024}$ by using the ratio of a simulation with better resolution, $P_{10, 1024}$, and an equally limited resolution as the reference, $P_{10, 512}$, $R_{{s}} = P_{10, 1024}/P_{10, 512}$, where $R_{{s}}$ is the mass resolution correction. The correction ranges from 5-14\% over the scales of interest and it increases with redshift \cite{bolton17}. These models are shown in Table I in \cite{Irsic2023}.

Furthermore, the finite box size naturally introduces a truncation of the flux power spectrum at large scales but also affects the small scales. This systematic is corrected with the box size correction from the ratio $P_{40, 2048}/P_{20, 1024}$. As shown in \cite{Irsic2023}, the correction at the low-$k$ modes is at the level of 3-5\%. We further check that the effect of mean flux rescaling in post-processing on the box size correction is $\leq$ 0.5\%. Therefore, we correct our models at fixed mean flux corresponding to ${\tau_{\mathrm{eff}}}^{\mathrm{B}}$.

\subsection{Patchy correction} \label{patchy}
Cosmic reionization is not homogeneous but occurs in three stages: the formation of H II bubbles, their overlap due to increased number of ionizing sources, and the disappearance of the last remaining neutral hydrogen  regions as a uniform UV background forms. As pointed out in previous work that has used the same suite of simulations (e.g. \cite{Molaro2021, Irsic2023}), the non-uniform nature of the reionization process leaves an imprint on the Lyman-$\alpha$ forest  at both large and small scales. The former consists of an enhancement of power at scales $k$ $<$ 0.03 km$^{-1}$s more noticeable at higher redshifts as well as thermal histories with an earlier  end of reionization. This can be understood through the temperature dependence of the recombination coefficient already mentioned in Subsection~\ref{flux_models}, which means that fluctuations in the gas temperature contribute  to the variations of the neutral hydrogen fraction. At small scales, patchy reionization causes suppression in the power spectrum mainly through thermal broadening and differences in peculiar velocity for regions that reionized recently \cite{Molaro2021, Nabendu2024}. We account for this extra power by computing the correction tabulated in \cite{Molaro2021} following \cite{Irsic2023}.

\begin{figure*}[hbtp!] 
    \centering
    \begin{minipage}[t]{0.38\textwidth} 
        \centering
        \vspace{30mm} 
        \textbf{$k$-cross Fold Validation}
        \includegraphics[width=1\textwidth]{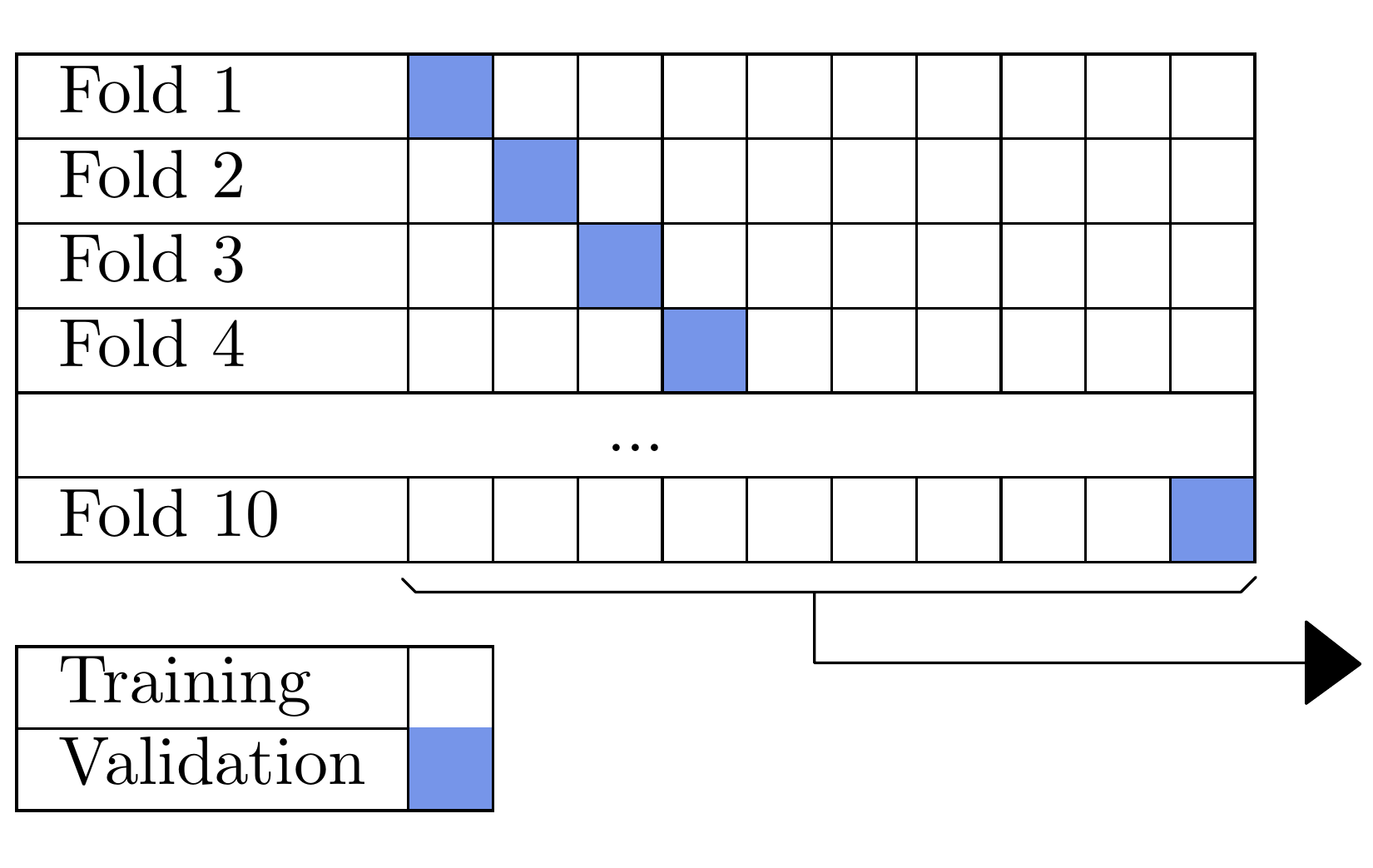} 
    \end{minipage} \hfill
     \hspace{0.0002\textwidth} 
    \begin{minipage}[t]{0.6\textwidth} 
        \centering
        \vspace{0mm}
        \includegraphics[width=1\textwidth,trim={2cm 11.7cm 0cm 0},clip]{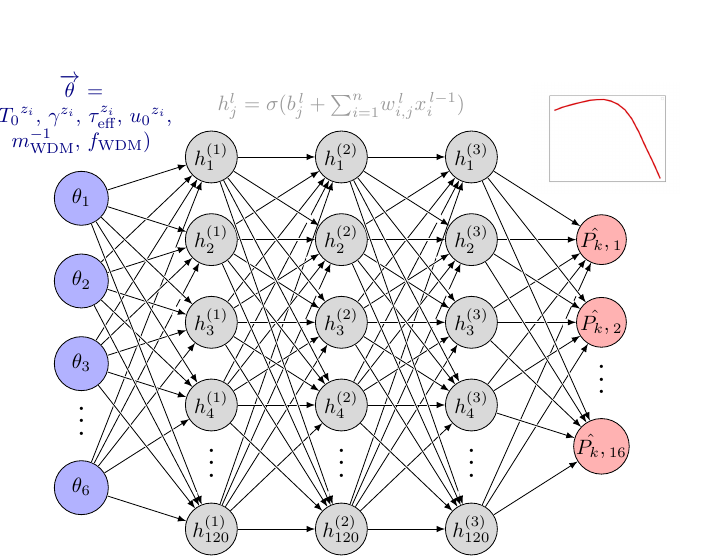} 
    \end{minipage}
    \caption{Illustration of the emulator building process used in this work. \textit{Left}: $k$-fold cross-validation. \textit{Right}: Neural network trained on each fold $i$ $\in$ [1,10]. We first extract the training set from the total input and output data to perform a $k$-fold cross-validation as shown on the left. Each fold is divided into 10 subsets, 9 are used to train the neural network model depicted on the right, while the remaining one is used as a validation set to monitor the convergence. We feed the cosmological model parameters $\overrightarrow{\theta}$ in the input blue layer. The hidden layers are shown in gray. For hidden neuron $j$ in layer $l$ with $n$ = 6 input parameters, the inputs to each node are combined in the weighted linear combination ${h^{i}}_{j}$ shown on top in gray, where $\overrightarrow{b}$ is the bias vector and $\textbf{w}$ is the weight matrix, such that $\phi$ = [$\overrightarrow{b}$,  $\textbf{w}$], $x^{\,l-1}$ are the outputs from the previous layer, and $\sigma$ is the so termed activation function, which permits non linear relations between the input and the output.
    This step results in 10 models validated on different subsets of the training data. Then, we feed the unseen test data into each model and average the predictions across all the emulators. The output are the power spectrum values in 16 $k$-bins shown in red, corresponding to the output layer. Note that this is repeated for the 3 redshift bins, $z$ = 4.2, 4.6, 5.0, of the data.}
    \label{nn}
\end{figure*}

\section{Statistical measures of the flux distribution} \label{section3}
The flux power spectrum models are defined on a six-dimensional grid per redshift bin, for each combination of model parameters (${T_{0}}^{z_{i}}$, ${\gamma}^{z_{i}}$, ${\tau_{\mathrm{eff}}^{z_{i}}}$, ${u_{0}}^{z_{i}}$, ${\mWDM}^{-1}$, $\fWDM$) =  $\overrightarrow{\theta}$. Comparing power spectra derived from observations and simulations enables us to constrain the free parameters of the model, $\overrightarrow{\theta}$, a process typically conducted within a Bayesian Monte Carlo Markov Chain (MCMC) framework.
In this study, we utilize the \texttt{Cobaya} code \cite{Cobaya} to sample the Gaussian multivariate likelihood function from \cite{Boera2019} at each redshift. However, our approach differs from previous works in the interpolation scheme used to generate simulated power spectra at each grid point. Specifically, we employ a machine learning model capable of predicting the power spectra within a fraction of a second on a single CPU.

\subsection{Neural network emulator} \label{emulator}
Given that $\overrightarrow{\theta}$ $\in$ ${\mathbb{R}}^{6}$, evaluating the likelihood at each grid point is limited by the computation cost of interpolating in such a high-dimensional space. We circumvent this problem by employing a neural network. This neural network maps the input parameters $\theta$ to the output parameters $P_{\text{F}}(k_m)$, in the form $P_{\text{F}}(k_m)$ = $f$[ $\overrightarrow{\theta}$,  $\phi$], where $\phi$ are the hyperparameters that describe the relationship between input and output, and $k_m$ is a finite set of wavenumbers matching the ones reported in the measurements of \cite{Boera2019}.

\begin{figure*}[htbp!]
\includegraphics[width=\linewidth]{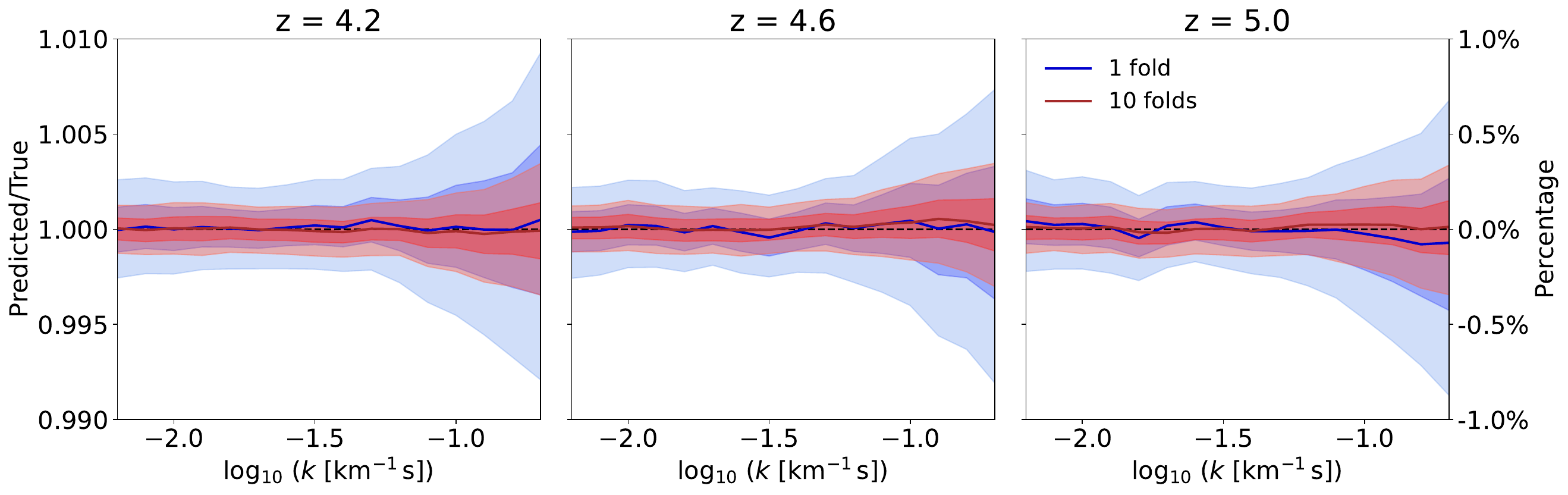}
\caption{Characterising the performance of the neural network emulator with the predicted versus true power spectra ratio distribution on the test set before and after performing $k$-fold cross-validation. Lighter blue and red regions correspond to the 95\% confidence intervals for one fold (before cross-validation) and for ten folds (after cross-validation), respectively. The same applies to the darker shaded regions, which correspond to the 68\% confidence interval. The median of the distribution is shown as well as a solid line. Perfect recovery  is indicated by the gray dashed line.} \label{error}
\end{figure*}

As an initial framework, we adopt the model architecture proposed by \cite{Molaro23} and make two key observations. First, the original work assessed the emulation error using $k$-fold cross-validation on the training set itself. This approach likely leads to overfitting, reducing the model’s ability to generalize to unseen data, which is precisely the role of an independent test set. Secondly, we can enhance the performance of the emulator by using a combination of baseline models, also known as ensemble learning. In particular, we consider averaging the output from a range of models trained using $k$-fold cross-validation. 

We begin by randomly splitting the dataset into a 90\% training set and a 10\% test set, with the validation set excluded within the training subset. We then divide the training set into $k$ subsets, and perform the $k$-fold cross-validation. In this way, each neural network is trained on $k$-1 folds of the training data, with the remaining one being used as validation. We use $k$ = 10, therefore repeating the process iteratively 10 times. This results in 10 neural network emulators, each validated on a different subset of the mock spectra. We build our neural network model using \texttt{PyTorch} in Python, which enables GPU execution.
Before training, we apply min-max normalization to rescale the input data, ensuring that $\overrightarrow{\theta}$ $\in$ [0,1], while the output (labels) undergo a logarithmic transformation. These preprocessing steps mitigate data variability, thereby enabling the network to learn more efficiently. We perform cross-validation for several hyperparameter combinations and, after monitoring the convergence of the loss function, we settle on a neural network with [120,120,120] units and a Rectified Linear Unit activation function (see Figure~\ref{nn}). We further find that the minimum of the validation loss is generally achieved after $\Delta t$ = 500 epochs. However, by allowing for early stopping, we note that the lowest redshift bin completes training earlier, whereas higher redshift bins do not meet the early stopping criteria.  We further optimize training by allowing the initial learning rate to decrease from an initial value of $10^{-3}$, to a minimum value of $10^{-6}$ if the validation loss does not improve for $\Delta t$ = 30 epochs. We choose the \texttt{Adam} optimization algorithm and a batch size = 18. Even though this batch size seems small compared to the total data size, we check that larger batches are not able to achieve as small losses for even larger number of epochs. 

This process of hyperparameter tuning is therefore done within cross-validation, where each neural network is validated using a different subset of the original training set, while the test set remains unseen. The building process of the emulator is shown in Figure~\ref{nn}.

In this way, we end up with 10 neural networks per redshift bin at our disposal. To obtain the final prediction, we average the prediction extracted from each model using the arithmetic mean. Since each model was trained in parallel and on a different training set in cross-validation, the combination of these models will give a better average performance than each model individually. We also use this method to estimate the combined emulator's error, as shown in Figure~\ref{error}. This shows that the error from the emulator is significantly smaller than the error on the data shown in Figure~\ref{fig:models} and the one used in \cite{Molaro23} without model ensembling and a separate test set.

\section{Results} \label{section4}

\begin{figure}[hbtp!] 
\includegraphics[width=\linewidth]{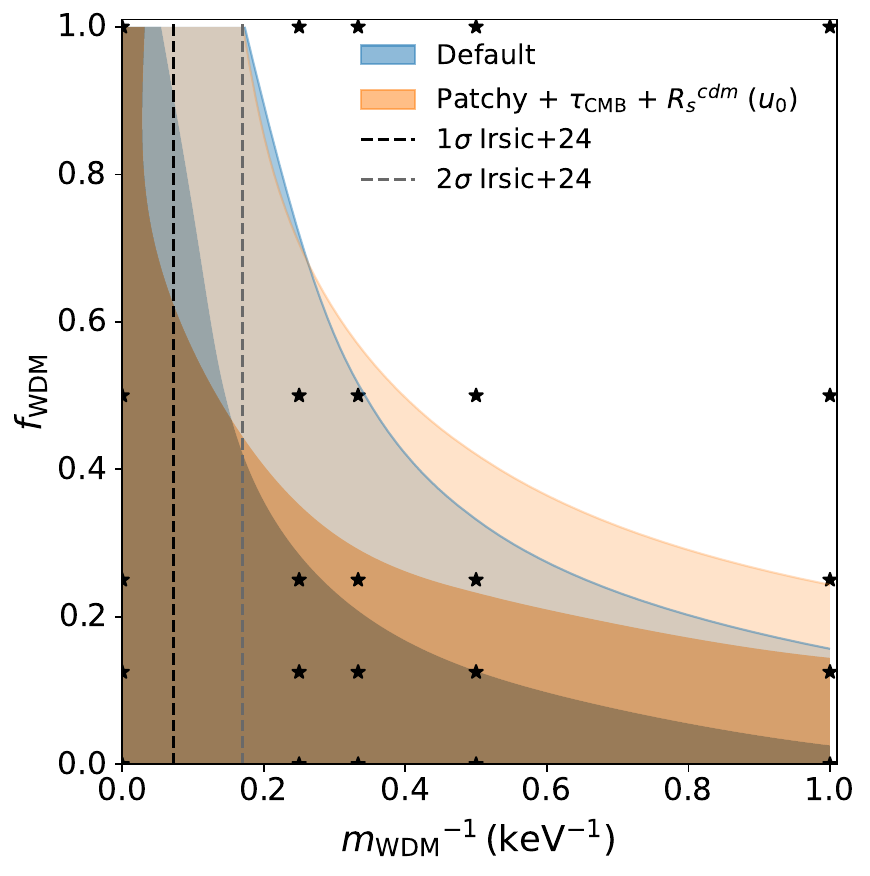}
\caption{The 2D posterior distribution in the $\fWDM$ and $\mWDM^{-1}$ plane for our default analysis (gaussian priors on $T_{0}$) and for patchy + $\tau_{\mathrm{CMB}}$ + ${R_{s}}^{\mathrm{cdm}}$ ($u_{0}$). Vertical black and gray dashed lines correspond to the 1$\sigma$ and 2$\sigma$ constraints from \cite{Irsic2023}. The grid of simulations is also shown as the starred black points.} \label{main}
\end{figure}

\begin{figure*}[hbtp!]
\includegraphics[width=0.8\linewidth]{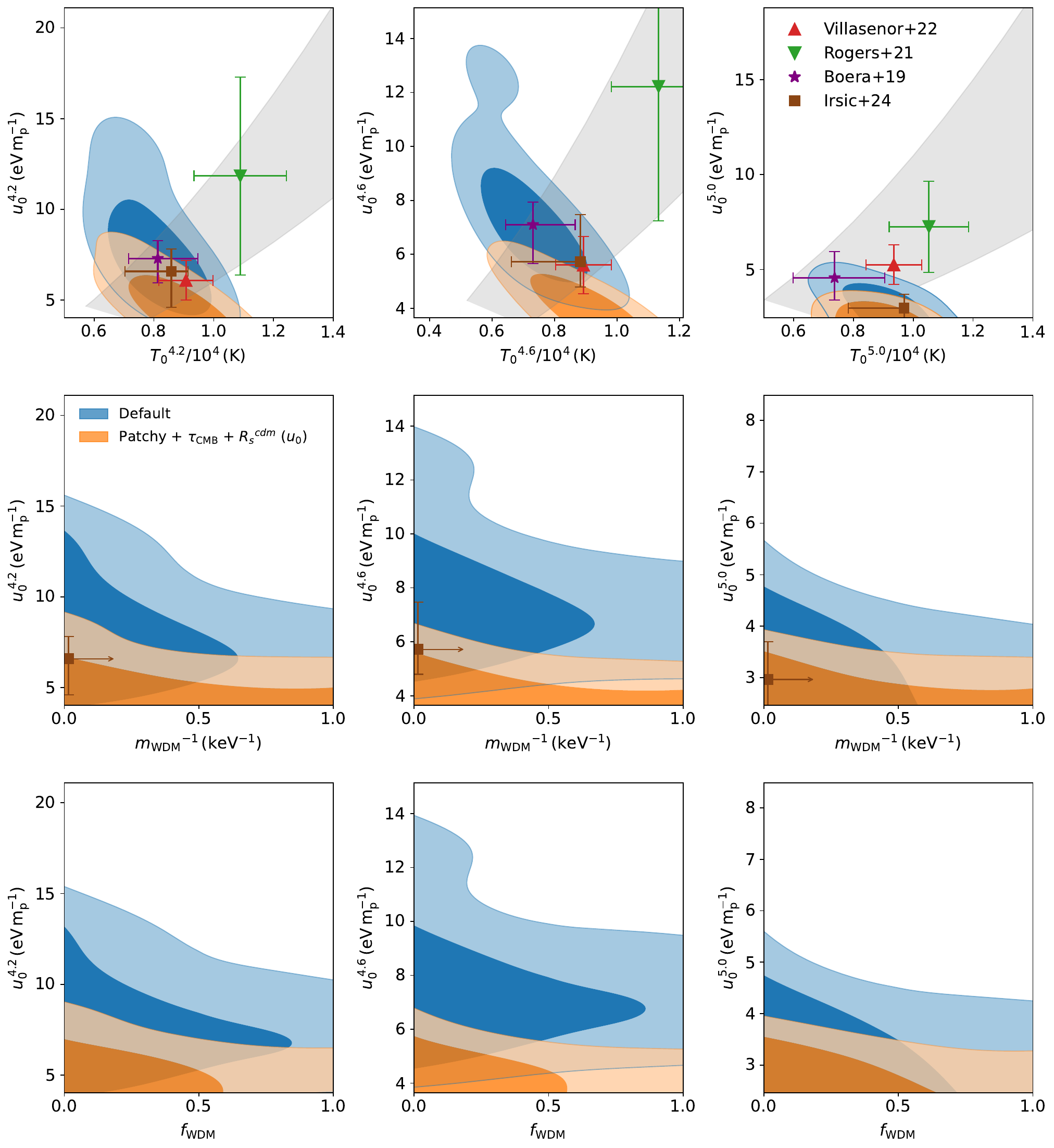}
\caption{2D posterior distributions for thermal parameters and dark matter parameters for the two analysis shown in Figure~\ref{main}. The analysis shown are corrected for resolution, ${R_{s}}^{\mathrm{cdm}}$, \textit{a posteriori}. The gray band represents the prior band imposed in the $u_{0}-T_{0}$ plane. Best-fit values for the same data-sets are also shown shown in red (\cite{villasenor22}), green (\cite{Rogers2021}), purple (\cite{Boera2019}), and blue (\cite{Irsic2023}). In the middle row, we show the 2$\sigma$ upper limit on ${\mWDM}^{-1}$ from \cite{Irsic2023} slighly shifted to the right for illustrative purposes. } \label{main_u0}
\end{figure*}

The results of the main analyses for thermal and dark matter parameters are shown in Figure~\ref{main} and Figure~\ref{main_u0}. We employ a uniform prior on ${\mWDM}^{-1}$, specifically $\mathcal{U} \sim [0, 1]$ ${\mathrm{keV}^{-1}}$, after verifying that the emulator's predictions are unreliable in extrapolated regions of the parameter space. Similarly, we impose a uniform prior on ${\fWDM}$, $\mathcal{U} \sim [0, 1]$. 
Figure~\ref{main} shows the 1 and 2$\sigma$ recovered posterior projected in the (${\mWDM}^{-1}-\fWDM$) plane. The shape of the contours matches the expected degeneracy between the two parameters, where the data allows for a very light thermal relic as long as its abundance is small (${\mWDM}^{-1}$ $\rightarrow$ 1), or a heavy thermal relic with large abundance ($\fWDM$ $\rightarrow$ 1), with CDM at (${\mWDM}^{-1}$, $\fWDM$) = (0,0) in this parameterization.
This degeneracy is well-fitted by a power law of the form  $\fWDM = A \times (1 \mathrm{keV}/\mWDM)^b$. For the 2$\sigma$ contours of the default analysis, we find $A$ = 0.14 $\pm$ 0.0007 for the normalization, and $b$ = -1.1 $\pm$ 0.0033 for the index parameter. 
We further identify a combination of  $\fWDM$ and ${{\mWDM}^{-1}}$ that is perpendicular to the above degeneracy axis, and is best-constrained by the data, through the parameter $W_{\rm WDM}$ $\approx$ $\fWDM$ (1 keV/$\mWDM$)$^{3.4}$. The 1$\sigma$ models in the 2D posterior of Figure~\ref{main} result in the upper bound $W_{\rm WDM}$ $<$ 0.27 at 1$\sigma$.   

\begin{figure}[hbtp!]
\includegraphics[width=\linewidth]{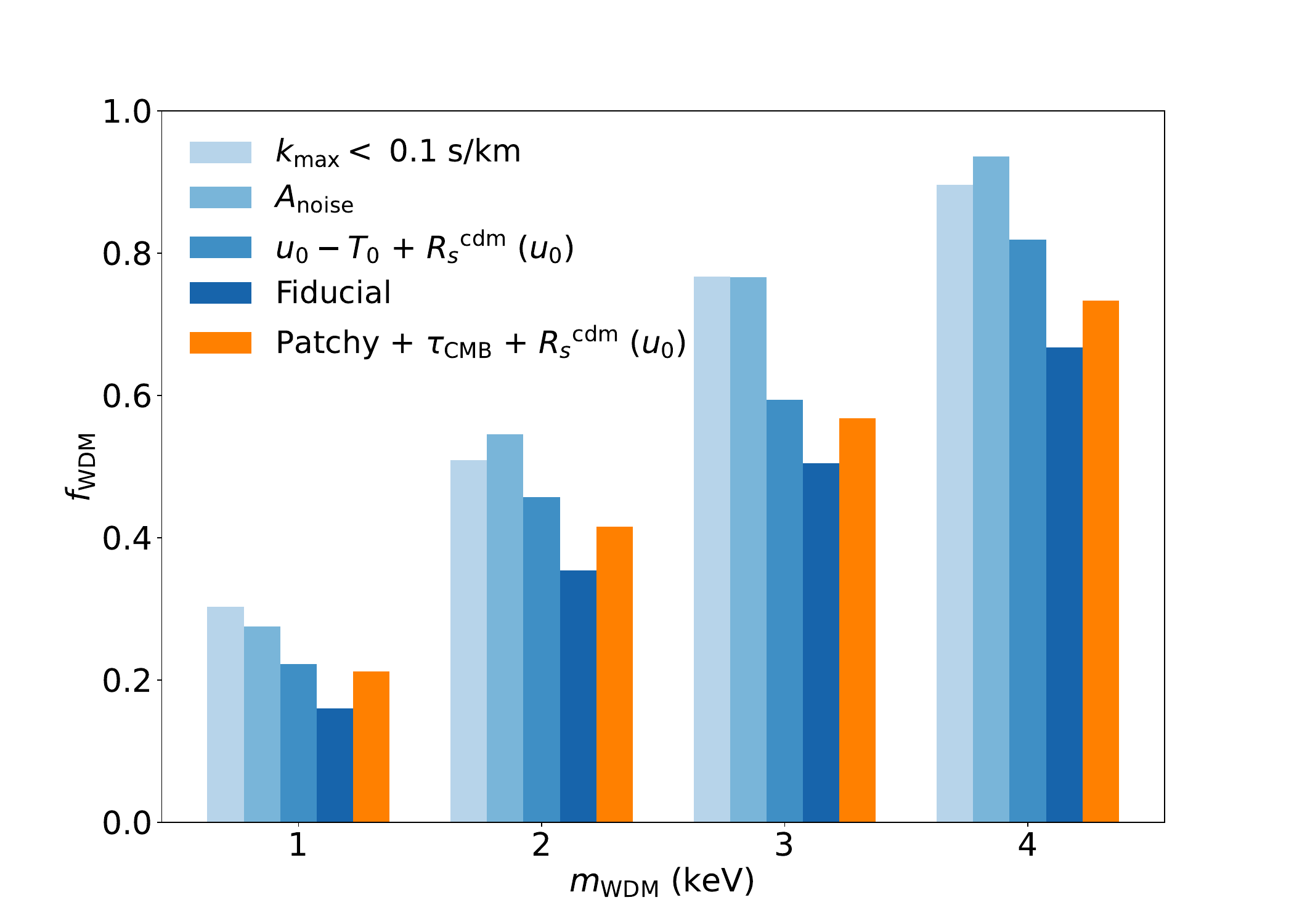}
\caption{Comparison of 2$\sigma$ upper bounds on $\fWDM$ at fixed $\mWDM$ bins.}  \label{2sig}
\end{figure}

\begin{figure*}[hbtp!]
\includegraphics[width=\linewidth]{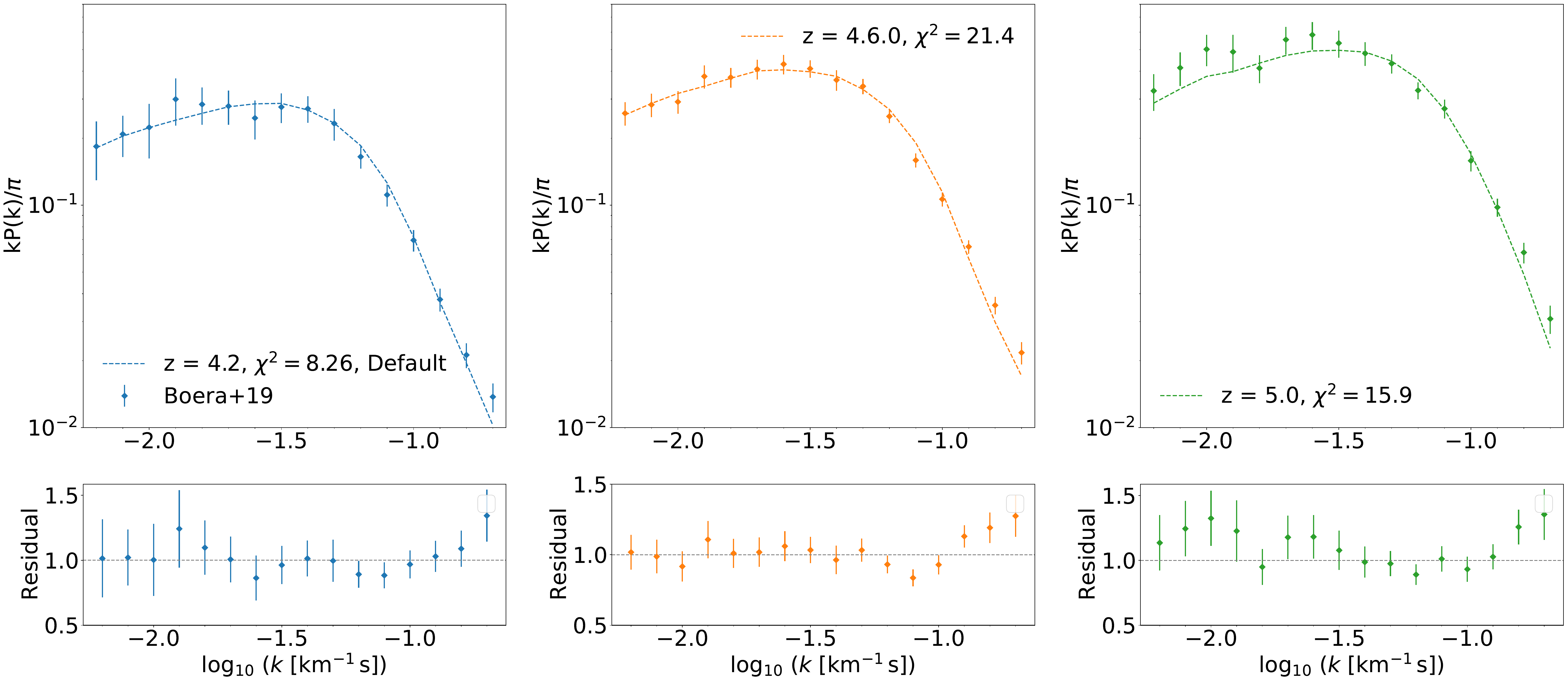}
\caption{Best-fit plot with individual ${\chi}^{2}$ for each redshift bin for our default analysis (Gaussian $T_{0}$ priors + ${R_{s}}^{\mathrm{cdm}}$) with four redshift dependent parameters ($T_{0}$, $\gamma$, $\tau_{\mathrm{eff}}$, $u_{0}$) and fixed dark matter parameters, ${\mWDM}^{-1}$ and $\fWDM$. The combined weighted $\chi^2$ is 45.6/34.} \label{fit}
\end{figure*}

\begin{figure}[hbtp!]
\includegraphics[width=\linewidth]{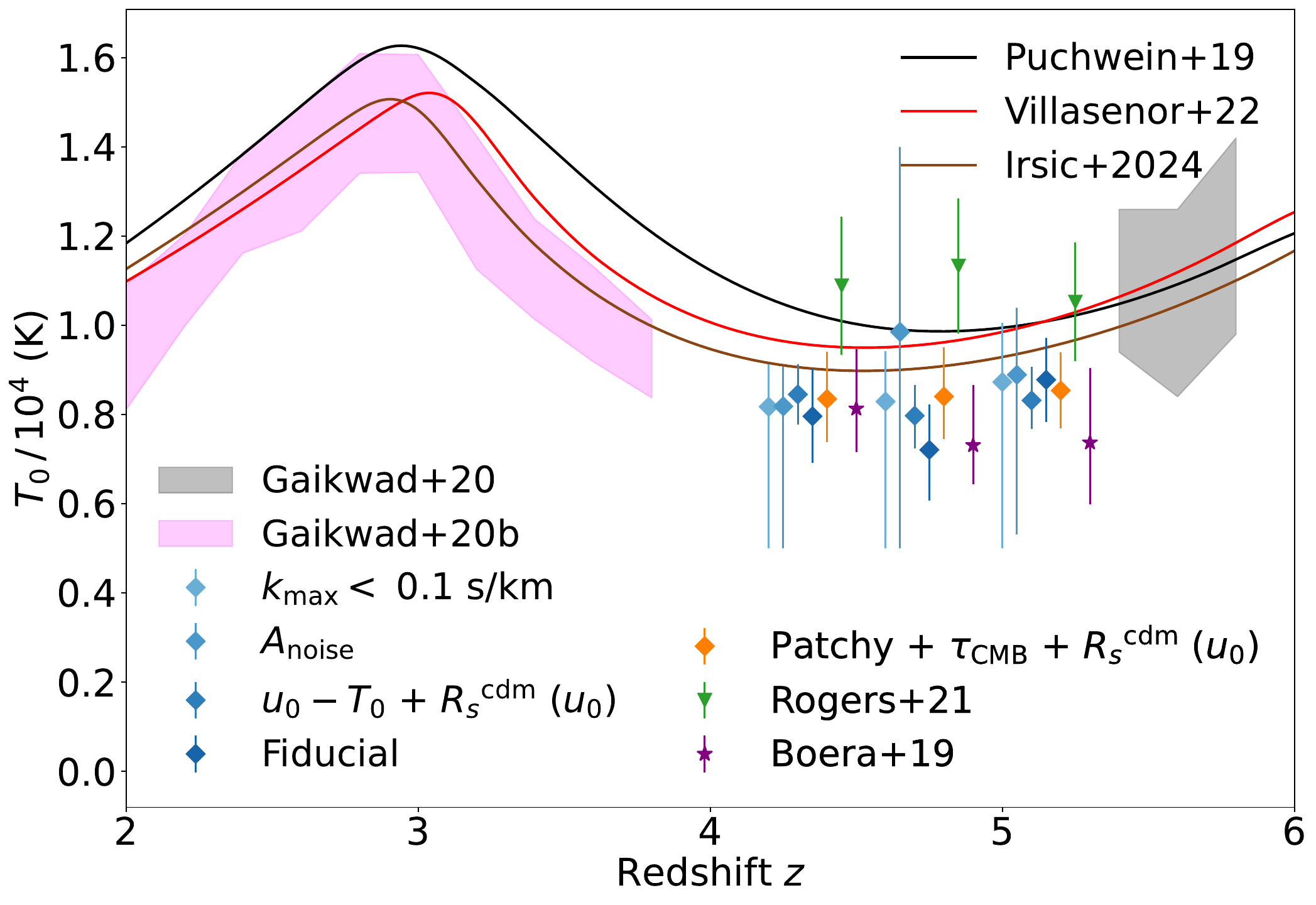}

\caption{Marginalized mean of the posteriors and corresponding error bars for $T_{0}$ ($z$) with other thermal evolution constraints. We show in different tones of blue the inferred temperature and mean density for a variety of analyses. The gray and pink bands show independent measurements from \cite{Gaikwad20} and \cite{Gaikwad21}. The black line show the predicted evolution from our simulated models (\cite{Puchwein19}). Inferred values from other works (\cite{Boera2019}, \cite{Rogers2021}, \cite{Villasenor2021} and \cite{Irsic2023}) are also shown. We have shifted the redshift axis for each new measurement for visual clarity.}  \label{t0evol}
\end{figure}

The default analysis uses gaussian priors on ${T_{0}}^{z_{i}}$ centered on fitted ${T_{0}}^{z_{i}}$ measurements from observations at low-$z$ (\cite{Gaikwad20}) and high-$z$  (\cite{Gaikwad21}), following \cite{Irsic2023}. The second analysis (orange) represents the model that incorporates all the physical considerations in addition to the default priors. These include a thermal-dependent resolution correction, patchy reionization correction and $u_{0}-T_{0}$ informed through the measurement of $\tau_{\mathrm{CMB}}$ from Planck \citep{planck18}. The following sections outline each correction and the resulting $\mWDM$ and $\fWDM$ constraints from the analyses, which are summarized in Figure~\ref{2sig}. For the two main models in Figure~\ref{main}, we show their fit to the data in Figure~\ref{fit}. We further show that the predicted IGM thermal evolution is in agreement with independent numerical and observational predictions in Figures~\ref{t0evol} and \ref{u0evol}. A summary of the results from all the analyses, including thermal and dark matter parameters along with their best-fit $\chi^2$ values, is presented in Table~\ref{results}.



\subsection{Thermal priors}
To effectively constrain the (${\mWDM}^{-1}-\fWDM$) plane, we need to marginalize over the two astrophysical effects that impact the small scale power spectrum apart from dark matter free-streaming.  

The thermal histories within our hydrodynamical simulations were constructed to bracket observational constraints on the thermal state of the IGM and effective optical depth evolution. These models can provide a useful prior in the $(u_0-T_0)$ plane, corresponding to the gray band in the 2D contours in the top row of Figure~\ref{main_u0}, as was done in \cite{Irsic2023}.

Alternatively, as proposed by \cite{Irsic2023}, one can impose a prior on the thermal state of the IGM represented by $T_{0}$, based on observational data and therefore independent of the specific physical models used in our analysis. We apply these priors in our default analysis shown in Figure~\ref{main_u0}. 

Finally, one can impose informed thermal priors through measurements of the Thomson scattering optical depth, denoted as $\tau_{\mathrm{CMB}}$, derived from Cosmic Microwave Background (CMB) data. This parameter offers a means to assess the IGM thermal history by quantifying the integrated scattering rate of photons by electrons over time, a quantity  which is inherently dependent on the electron number density \cite{Puchwein2023}. Consequently, for each thermal history model, there exists a mapping between the integrated and instantaneous thermal history, which is parameterized by ${u_{0}}^{z_{i}}$ and ${T_{0}}^{z_{i}}$, and the electron scattering optical depth. We use this mapping to interpolate $\tau_{\mathrm{CMB}}$ for each sampled combination (${u_{0}}^{z_{i}}$, ${T_{0}}^{z_{i}}$). At the likelihood level, we introduce an additional Gaussian term for $\tau_{\mathrm{CMB}}$, centered on $\tau_{\mathrm{CMB}}$ = 0.054 $\pm$ 0.007 as inferred by Planck \cite{planck18}. While the derived $\tau_{\mathrm{CMB}}$ is heavily dependent on our simulated models, the prior effectively constrains the thermal parameters in a manner that ensures the preferred cosmology is consistent with CMB measurements of this parameter. 

We find similar results to \cite{Irsic2023} in that the contour in the ($u_{0}-T_{0}$) plane for the default analysis expands along the anti-correlation direction, with slightly colder models allowed within 1$\sigma$ at $z$ = 4.2 and 4.6, as shown by the blue contours in the upper row of Figure~\ref{main_u0}.  This then results in a slight upward shift in the ($u_{0}-{\mWDM}^{-1}$) plane in the middle row, with colder models showing less pressure smoothing. The effect of different thermal priors does not have a noticeable effect in the (${\mWDM}^{-1}-\fWDM$) plane (bottom row); therefore, the 2$\sigma$ constraints for pure WDM (see Table~\ref{results}) are also very similar. The addition of $\tau_{\mathrm{CMB}}$ only slightly shrinks the posterior in the thermal parameter space at $z$ = 4.2, resulting in smoother 1$\sigma$ contours, but the inferred parameters are practically the same. For mixed DM models, we find that $\fWDM$ $\lesssim$ 16\%, 35\%, 50\% and 67\% for fixed $\mWDM$ = [1,2,3,4] keV, respectively, where the choice of different thermal priors only changes these constraints by $\leq$ 5\%. 

\subsection{Effect of small scale cut} \label{kmax}

The previous data sets used to infer the WDM constraints extend
to $k$ $\approx$ 0.1 s/km \cite{Viel13wdm, Irsic17b,Irsic2023}. We check whether we reproduce their results on pure WDM thermal relics by performing
an analysis with both uninformative and default thermal priors and without the last 3 $k$-bins in the data. These scales are key to disentangle thermal effects from dark matter free-streaming, meaning that without them, constraints in this parameter space will be dominated by the priors. The best-fit model for the default analysis with the small scale cut precisely recovers ${T_{0}}^{z_{i}}$, within the Gaussian prior's peak. The case for uninformed priors expands along the $u_{0}-T_{0}$ degeneracy axes, effectively showing no sensitivity to these parameters at the scales considered. The recovered $\chi^2$ when including or not including thermal priors are 12.3/20 and 11.6/20, respectively. 
We further recover weaker constraints for pure $\mWDM < $ 3.67 and $\mWDM < $ 3.82 keV (2$\sigma$) for each analysis, consistent with \cite{Villasenor2023, Irsic2023}. In terms of mixed dark matter models, the scale cut weakens the allowed $\fWDM$ by 15\%, with a stronger effect for $\mWDM$ = 3 keV, where the difference is $\approx$ 30\% as shown in  Figure~\ref{2sig}.

\subsection{Effect of noise}
The small scale power in the \lya power spectrum includes contributions from metal lines, instrumental resolution, and potentially underestimated instrumental noise.
As mentioned in Section~\ref{section1}, we correct for instrumental resolution and pixel size of the data following \cite{Boera2019}. Furthermore, we use the analysis results of \cite{Boera2019} that corrects for the effect of the intervening metal lines.
However, to address instrumental noise, \cite{Boera2019} estimated the "noiseless" power spectrum by subtracting the white noise power corresponding to the mean of the noise distribution, which dominates the measurement at high $k$ values. 
\cite{Irsic2023} fitted the noise data for each redshift bin, as estimated by \cite{Boera2019}, to a log-normal distribution. To assess whether the noise flux power may be  underestimated in the data, \cite{Irsic2023} added an additional term to the theoretical model, which accounts for deviations from the mean of the noise distribution.  We model the noise in the same way, including an extra parameter with log-normal priors, ${A_{\mathrm{noise}}^{z_{i}}}$, alongside the power spectra predicted by the emulator as given by Eq. 4 in \cite{Irsic2023}. This analysis excludes pure thermal relics with $\mWDM$ $<$ 3.82 keV (2$\sigma$), in agreement with \cite{Irsic2023}. 

\begin{figure}[hbtp!]
\includegraphics[width=\linewidth]{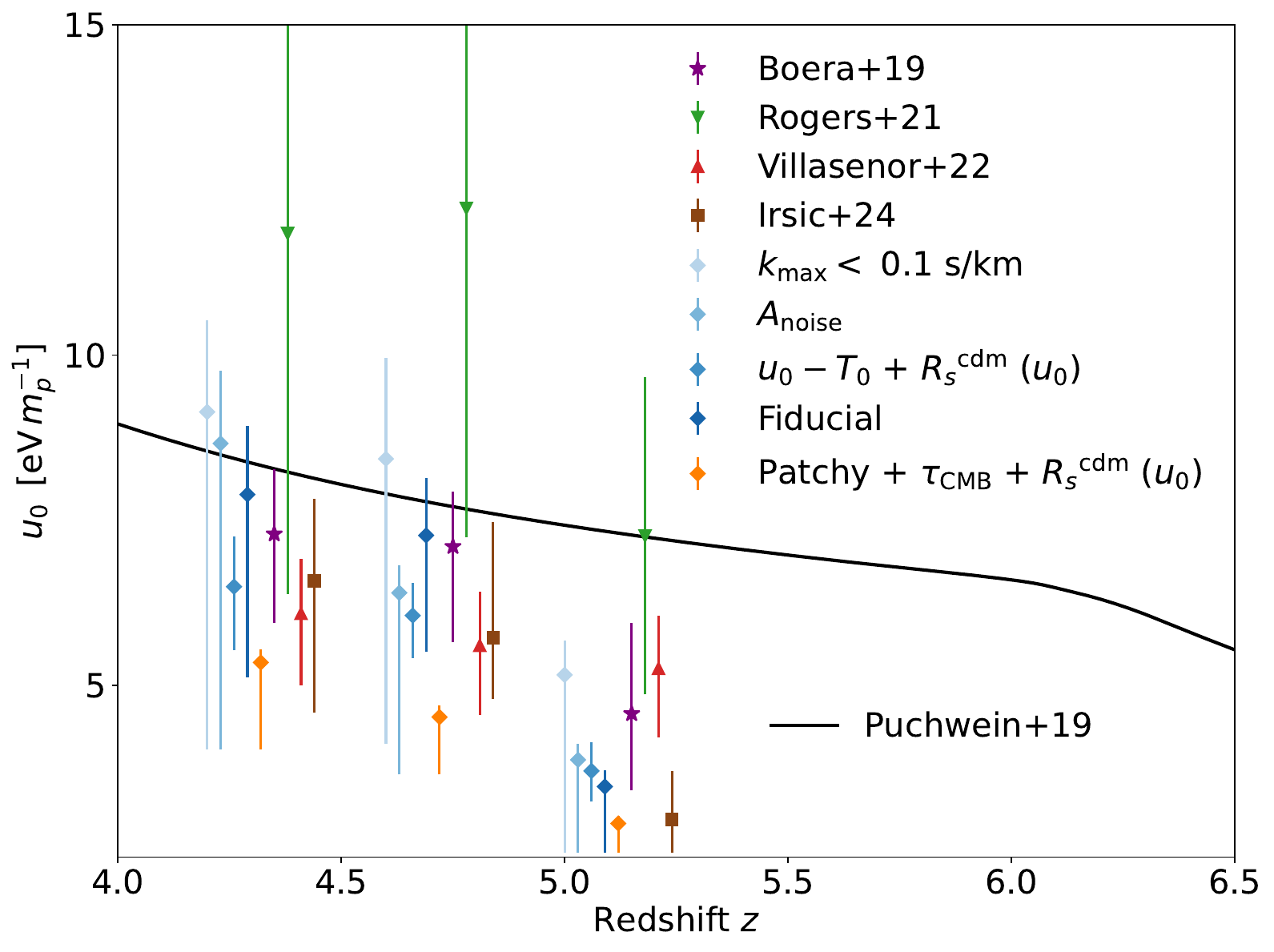}
\caption{Marginalized mean of the posteriors and corresponding error bars for $u_{0}$($z$) with other inferred constraints at the same redshift bins from \cite{Boera2019}, \cite{Rogers2021}, \cite{Villasenor2021} and \cite{Irsic2023}).} \label{u0evol}
\end{figure}

Moreover, we recover the same behavior for ${A_{\mathrm{noise}}^{z_{i}}}$, which remains largely insensitive to ${\mWDM}^{-1}$ (that is, the free-streaming effect), while exhibiting a correlation with the thermal parameters ${T_{0}}^{z_{i}}$ and ${u_{0}}^{z_{i}}$. This correlation with the IGM temperature favors hotter models (see Table~\ref{results}), whereas its anti-correlation with ${u_{0}}^{z_{i}}$ suggests a preference for later reionization, implying smaller heat injection.
Importantly, we obtain mean values for the posteriors of ${A_{\mathrm{noise}}^{z=4.2}} = {0.72}^{+0.98}_{-0.34}$,
${A_{\mathrm{noise}}^{z=4.6}} = {1.19}^{+1.61}_{-0.85}$,
and ${A_{\mathrm{noise}}^{z=5.0}} = {1.19}^{+1.65}_{-0.78}$,
suggesting that the noise in the data might indeed be underestimated. Compared to \cite{Irsic2023}, the 1$\sigma$ error is broader, primarily due to the large uncertainty in ${T_{0}}^{z_{i}}$, which allows for a very warm IGM temperature, thereby weakening constraints on the WDM mass. The preference towards less heat injected (lower $u_{0}$) is particularly noticeable at $z$ = 4.6, as shown in Figures~\ref{t0evol} and \ref{u0evol}.

Regarding CWDM constraints, we find that the upper bounds on $\fWDM$ decrease by up to 25\% (see Figure~\ref{2sig}), yielding very similar results to those obtained in the scale-cut analysis as described in Subsection~\ref{kmax}.

\subsection{Effect of mass resolution}
Our emulator provides an averaged prediction of the absolute binned flux power spectrum, as described in Section~\ref{emulator}. Assuming that the box size and mass resolution corrections are independent of the parameters in our flux models, we apply these corrections \textit{a posteriori}, that is, after obtaining the predicted flux power spectrum from the emulator. This corresponds to the default analysis resulting in the fit shown in Figure~\ref{fit}.

However, this simplified method does not account for the fact that the mass resolution correction depends on thermal parameters. The missing small-scale perturbations in the flux power spectrum are influenced by the initial perturbations in this regime. These perturbations contribute to the formation of small-scale structures, which are smoothed out in models with more cumulative injected heat (higher $u_{0}$). As a result, models where the hydrodynamic response naturally suppresses small-scale perturbations tend to be more numerically converged, leading to a smaller resolution correction. This effect was explored in \cite{Irsic2023}, which allowed this work to impose even tighter constraints on dark matter free-streaming.
We re-compute the models and find that scenarios with earlier reionization require a correction that is about 5\% smaller than that required for colder models. This refined resolution correction is denoted as ${R_{s}}^{\mathrm{cdm}}$ ($u_{0}$). In practice, we also apply this correction  \textit{a posteriori}, interpolating in  ${u_{0}}^{z_{{{i}}}}$ on the fly during sampling. 

We find that including the thermal dependence in the mass resolution tightens the posteriors in the thermal parameter space while broadening the contours in the dark matter parameter space, ultimately leading to weaker constraints on $\mWDM$ and $\fWDM$. This conclusion holds for the default Gaussian $T_{0}$ priors, and when including $u_{0}-T_{0}$ priors on top of the default prior, with the latter exhibiting only a slightly reduced constraining power in the high ${\mWDM}^{-1}$ regime due to the posteriors in $u_{0}-T_{0}$ being confined to the envelope of the simulations. For a pure WDM cosmology, we obtain a lower bound of $\mWDM >$ 4.66 keV for $\fWDM=$ 1 when accounting for the thermal dependence of mass resolution, similar to bounds found by \cite{Irsic2023}. The constraints on $\fWDM$ and fixed mass bins become weaker by 5-10\% (Figure~\ref{2sig}).

\subsection{Effect of patchy correction}
We account for the effect of patchy reionization on the default analysis by repeating the inference using a new set of emulators trained on models that include the patchy correction described in Section~\ref{patchy}. With default thermal priors, we find that the data tends to prefer cold models with very low $u_{0}$ values. This was observed in \cite{Irsic2023}, and occurs due to patchy-corrected models showing less structure at the small scales as a result of the peculiar velocity field structure \cite{Molaro2022}. 
These models therefore match a very late end of reionisation, eventually becoming unphysical. If we impose a $u_{0}-T_{0}$ prior, the posterior still shifts to lower values but within the envelope of simulations, leaving more room for dark matter free-streaming. The bounds on WDM now become $\mWDM$ $>$ 5.86 keV, with little change of $\leq$ 5\% on mixed models.

Adding the thermal-dependent resolution correction shifts the posterior in the other direction, since resolution correction increases flux power at small scales, increasing with decreasing heat injected during reionization. As a result, slightly warmer IGM temperatures are allowed with $u_{0}$, specially at high $z$, pushing the lower bound. As a result the constraints slightly weaken to $\mWDM$ $>$ 5.11 keV with the $u_{0}-T_{0}$ prior, and to a lesser extent ($\mWDM$ $>$ 5.39 keV) with the $\tau_{\mathrm{CMB}}$ prior. We show this behavior in orange for the analysis with thermal priors using  $\tau_{\mathrm{CMB}}$, patchy and a thermal-dependent resolution correction in Figure~\ref{main_u0}. We further note that the $\chi^2$ does not change noticeably compared to the default analysis when both patchy and ${R_{s}}^{\mathrm{cdm}}$ are modeled (see Table~\ref{results}). The CWDM constraints, as shown in Figure~\ref{2sig}, are stronger than the other analyses and comparable to the default one. From the 2$\sigma$ contour in Figure~\ref{main}, the patchy + $\tau_{\mathrm{CMB}}$ + ${R_{s}}^{\mathrm{cdm}}$ ($u_{0}$) orange model is more inflated than the fiducial model towards high ${\mWDM}^{-1}$, yielding slightly weaker constraints at the fixed mass bins shown in Figure~\ref{2sig}. In general, the difference with respect to other analyses is more noticeable in the intermediate mass bin, $\mWDM$ = 3 keV, corresponding to the region in Figure~\ref{main} where the degeneracy between $\fWDM$ and ${\mWDM}^{-1}$ is stronger. While these models incorporate physical effects applied \textit{a posteriori}, the resulting constraints remain unchanged, as the opposing effects of patchy reionization and resolution correction counterbalance each other. To the best of our knowledge, this approach represents the most comprehensive level of modeling. 
Investigating a cosmology-dependent resolution correction, alongside other effects including peculiar velocity \cite{Irsic2023}, will improve our understanding of the recovered bounds for each mass bin in the context of CWDM models. This is particularly crucial in the regime $k$ $>$ 0.1 km$^{-1}$s, where the sensitivity to dark matter free-streaming is the highest.

\begin{table*}[hbtp!]
\renewcommand{\arraystretch}{1.2} 
\setlength{\tabcolsep}{4pt}
\caption{Best-fit constraints on thermal and dark matter model parameters for different analysis at $z$ = 4.6. Default corresponds to using Gaussian $T_{0}$ priors and mass resolution correction ${R_{s}}^{\mathrm{cdm}}$ except for the thermal resolution dependent analysis where ${R_{s}}^{\mathrm{cdm}}$ ($u_{0}$) is applied instead. The remaining analysis are additive on top of the default.} \label{results}
\begin{tabular}{lcccccc}
\hline
Analysis & $T_{0}$  & $u_{0}$ & $\mWDM$ (2$\sigma$) & $A_{\mathrm{noise}}$ & $\chi^{2}$/dof  
\\ 
& [\,$10^{4}$ K] & [eV\,${\mathrm{m_p}}^{-1}$] & [keV] ($\fWDM$=1)  &  
\\
\\ \hline
Default & ${0.714}^{+0.113}_{-0.099}$ & ${6.461}^{+2.10}_{-0.58}$ & $>$ 5.47 & -&45.6/34  \\ 
$\tau_{\mathrm{CMB}}$ & ${0.749}^{+0.066}_{-0.137}$ & ${6.362}^{+2.149}_{-0.331}$ & $>$ 5.52 & - &  47.6/34  
\\
 \hline
$u_{0}-T_{0}$  &  ${0.749}^{+0.101}_{-0.033}$ &  ${6.349}^{+0.612}_{-0.644}$ & $>$ 5.42 & -& 47.1/34    \\
\hline 
$k_{\mathrm{max}} = 0.1$ s/km & ${1.18}^{+0.22}_{-0.68}$ &  ${5.03}^{+4.93}_{-0.9}$ & $>$ 3.58 & -& 11.6/20 \\
\hline
$A_{\mathrm{noise}}$ & ${1.30}^{+0.1}_{-0.8}$ & ${3.87}^{+2.93}_{-0.221}$ & $>$ 3.82 & ${1.30}^{+0.1}_{-0.09}$ & 15.3/31 \\ 
\hline
$\tau_{\mathrm{CMB}}$ + ${R_{s}}^{\mathrm{cdm}}$($u_{0}$) & ${0.862}^{+0.021}_{-0.201}$ & $<$ $5.92$ & $>$ 5.22 &-& 43/34 \\ 
$u_{0}-T_{0}$ + ${R_{s}}^{\mathrm{cdm}}$($u_{0}$) & $0.893^{+0.107}_{-0.275}$ & $<$ 5.75 &  $>$ 4.66 & - & 39.2/34 \\ 
\hline 
patchy + $u_{0}-T_{0}$ & ${0.754}^{+0.068}_{-0.07}$ & $<$ 6.28 & $>$ 5.86 &-& 55.7/34\\ 
patchy + $u_{0}-T_{0}$ + ${R_{s}}^{\mathrm{cdm}}$ ($u_{0}$)& ${0.788}^{+0.043}_{-0.089}$ & $<$ 4.70 & $>$ 5.11 &-& 46/34\\ 
patchy + $\tau_{\mathrm{CMB}}$ + ${R_{s}}^{\mathrm{cdm}}$ ($u_{0}$) & $0.885^{+0.055}_{-0.1345}$ & ${3.69}^{+1.01}_{-0.94}$& $>$ 5.39 & - & 45.2/34\\ 
\hline 
\end{tabular}
\end{table*}

\subsection{Comparison to previous CWDM analyses}
The main results of this work are shown in Figure~\ref{2sig}. A thermal relic as light as $\mWDM$ = 1 keV is allowed by the data if it contributes $<$ 16\% at 2$\sigma$ to the total dark matter energy density, with increasing $\fWDM$ for increasing mass. The pure WDM limit in the default analysis results in a $\mWDM$ $>$ 5.47 keV (2$\sigma$), which is slightly weaker than the mass bound found by \cite{Irsic2023}. However, within our set of analyses we also recover the $\mWDM$ $>$ 5.86 keV (2$\sigma$) constraints (Table~\ref{results}). These lower bounds are highly sensitive to the modeling of $P_{\mathrm{F}}$ ($k$), including the corrections described in Subsection~\ref{Rs} and \ref{patchy}. Moreover, incorporating $\fWDM$ into the $\overrightarrow{\theta}$ model parameters requires sampling a higher-dimensional parameter space during MCMC compared to \cite{Irsic2023}. This introduces non-trivial degeneracies, particularly the one illustrated in Figure~\ref{main}, which requires a substantial number of samples to achieve convergence. Notably, the quoted limits have been derived by filtering the original samples from CWDM runs, in contrast to fixed mass bin analyses conducted for axions \cite{eROSITA, Calabrese25}. A similar degeneracy between the fraction and the axion mass has been observed also in previous work based on simulations of mixed dark matter \cite{Kobayashi17}, with updated constraints by \cite{Rogers2021} using the same Lyman-$\alpha$ data from \cite{Boera2019} as in this work. The degeneracy in the context of CWDM models, however, appears in the keV mass range.

The main two analyses carried out in the context of Lyman-$\alpha$ CWDM constraints are \cite{Boyarsky2009, Baur2017}. The former found constraints on thermal relics that are significantly weaker.
The constraints found in the main analysis of the former are weaker, allowing the smallest thermal-relic mass probed, $\mWDM$ = 1.1 keV, for $\fWDM$ $<$ 40\%, with the pure WDM limit $\mWDM \geq 1.7$ keV (95\% C.L.). These constraints are not competitive with the results presented in Figure~\ref{main} mainly due to the lower resolution of the data used (SDSS and the VISTA Hemisphere Survey with $k_{\mathrm{max}}$ = 0.03 km$^{-1}$s).

\cite{Baur2017} analysed BOSS DR9 data combined with XQ-100, HIRES, and MIKE, finding $\mWDM > 0.7$ keV for $\fWDM < 0.10$. Their $\mWDM$ bound falls outside our simulation grid in the ($\fWDM$ – $\mWDM^{-1}$) plane. Their power-law fit to the posterior in this plane, given by $\fWDM = A \times (1 \mathrm{keV}/\mWDM)^{b}$, yielded the same $A$ value but smaller $b$ compared to the fit described in Section~\ref{section4}. This yields an inflated posterior for $\fWDM \to 1$ (larger $\alpha$), leading to weaker WDM constraints, which declines more steeply, with $\fWDM \leq 0.17$ at $\mWDM = 1$ keV, almost identical to our constraints. Compared to \cite{Baur2017}, the work presented in this paper uses improved simulations that vary thermal history for all the CWDM models considered. The effects of thermal history are marginalized over a more physical and conservative prior range than in \cite{Baur2017}. This would weaken the resulting constraint on CWDM models, a point discussed in previous works \cite{Irsic2017b,Baur2017}. However, the improved resolution (higher $k_{\mathrm{max}}$) and treatment of systematics (e.g. noise, metal contamination) in \cite{Boera2019} counterbalances this effect, yielding slightly stronger constraints for these mixed models.


The results by \cite{Baur2017} show that, as ${\mWDM}^{-1}$ increases, the 1$\sigma$ contour in the ($\fWDM$ – $\mWDM^{-1}$) plane shrinks when using higher resolution data. This would in principle mean that there exists a light thermal relic for which the contour closes, meaning that Lyman-$\alpha$ forest data would be able to completely exclude CWDM models beyond a $\mWDM$ value. 

In our analysis, however, we use data that reaches to higher $k_{\mathrm{max}}$ than in \cite{Baur2017}. 
While the posterior in Figure~\ref{main} shrinks due to the small-scale information available from \cite{Boera2019} data, we expect the high $\mWDM^{-1}$ regime to achieve a constant plateau. This arises from the imprint that these light thermal relics ($\mWDM$ $<$ 1 keV) leave on the matter power spectrum, as discussed in Section~\ref{section2}. For large $\mWDM^{-1}$, suppression in the matter power spectrum shifts to lower $k$-modes. While the Lyman-$\alpha$ forest data is sensitive to the mildly non-linear scales, and therefore not to the free-streaming scale from these light thermal relics, the plateau reached by these mixed dark matter models shifts the overall amplitude of $P_{\mathrm{F}}(k)$. This effect is degenerate with other astrophysical effects, mainly with the mean IGM transmission, $\tau_{\mathrm{eff}}$. The difference between these flux models is therefore smaller than the error on the data, shown in Figure~\ref{fig:models}, meaning that the data will not be sensitive to them. Consequently, even if additional simulations were run for $\mWDM$ $>$ 1 keV, we expect the data to maintain the constraint on the fraction $\fWDM$ at the same level. This is likely true for the models where the half-mode scale is smaller than $0.05$ $\mathrm{Mpc}^{-1}$, at which point marginalising over other cosmological parameters (e.g. $A_s,n_s$) would become important.

\section{Conclusions}
\label{section5}
In this work, we present updated constraints on CWDM models using new 1D flux power spectrum measurements from the high redshift Lyman-$\alpha$ forest provided by \cite{Boera2019}. The statistical power of this data set is limited by cosmic variance as a result of the small number of QSO sightlines available. However, the number of sightlines has more than doubled compared to previous analyses using the same spectrographs, and the improvement in the systematic modeling provides access to scales a factor of $\approx$ 2 smaller than those probed in previous WDM and CWDM constraints. This enables direct measurement of the flux power spectrum at wavenumbers $k$ $\geq$ 10 $h\,$$\mathrm{Mpc}^{-1}$, precisely where free-streaming effects from the WDM component are expected to suppress structure formation in these hybrid models.

Our analysis indicates that a thermal relic with $\mWDM = 1$ keV is allowed if its contribution to the total dark matter density does not exceed 16\% (2$\sigma$). For higher mass bins ($\mWDM = 2, 3, 4$ keV), increasingly larger abundances are permitted, with $\fWDM = 0.35$, 0.50, and 0.67 at 2$\sigma$. These findings refine previous constraints from \cite{Boyarsky2009}, which were based on lower-resolution SDSS and VHS data, as well as a limited number of simulations for low $\mWDM$ values. Additionally, our results align closely with those of \cite{Baur2017} on CWDM in the low-$\mWDM$ regime explored by our simulations, with minor differences arising from variations in thermal history treatment,  simulation runs at lower $\mWDM$ grid points, and improved data from HIRES.

On the pure WDM side, we also obtain stronger constraints than previous studies (\cite{Boyarsky2009, Irsic17b, Villasenor2023}, due to the inclusion of higher $k$-bin information. These constraints remain consistent with those reported by \cite{Irsic2023}, although fiducial analysis results in slightly weaker lower bounds. This is primarily attributed to the extensive sampling required to explore the high-dimensional parameter space, particularly the degeneracy between $\fWDM$ and ${\mWDM^{-1}}$. However, in certain analyses, we derive even more stringent constraints. For instance, the patchy + $u_{0}-T_{0}$ model yields $\mWDM$ $>$ 5.86 keV compared to $\mWDM$ $>$ 5.10 keV in \cite{Irsic2023}. This suggests that incorporating additional physical effects may alleviate the strong degeneracies among sampled parameters. Specifically, patchy models consistently favor scenarios with reduced heat injection, which may, in turn, enhance the efficiency of likelihood sampling.

Beyond Lyman-$\alpha$ forest analyses, CWDM has also been constrained by combining BOSS data with Planck and estimates of the number of satellite galaxies \cite{Diamanti17}. Our constraint at the lowest mass bin probed in this analysis, $\mWDM$ = 1 keV, is 10\% stronger. More recently, \cite{Facchinetti25} demonstrated that the combination of CMB data and the predicted 21 cm signal power spectrum yields very weak constraints in the ($\fWDM-{\mWDM}^{-1}$) plane, with a lower bound of $\mWDM$ $>$ 1.8 keV ($\fWDM$ = 1), in contrast to the significantly stronger constraints from the Lyman-$\alpha$ forest given in this work. In particular, the main findings of this work can be summarized as:
\begin{itemize}
    \item We provide new updated constraints on CWDM cosmology, where a thermal relic with $\mWDM = 1$ keV is allowed for $\fWDM$ $<$ 0.16 (2$\sigma$). Higher mass bins are viable with increasing abundances, leading to a strong degeneracy between $\fWDM$ and ${\mWDM}^{-1}$ parameters. The 2$\sigma$ contour ($\fWDM-{\mWDM}^{-1}$) plane is well-described by $\fWDM = 0.14$ $(1 \mathrm{keV}/\mWDM)^{-1.1}$. 

    \item We update previous Bayesian-inference framework for 1D flux power spectrum Lyman-$\alpha$ forest analysis by integrating a neural network emulator. Model ensembling via $k$-fold cross-validation reduces the emulator error to $<$ 0.5\% at 2$\sigma$ across all $k$-bins.
    \item We recover a thermal state and evolution of the IGM consistent with previous Lyman-$\alpha$ forest analyses (\cite{Boera2019, Villasenor2023, Rogers2021}), and external measurements (\cite{Gaikwad20, Gaikwad21}, with very mild variations in $T_{0}$ and $u_{0}$ due to the strong degeneracy between these two parameters. The mean IGM transmission is also in agreement with independent measurements from e.g. \cite{Becker13, Eilers18, Bosman22}.
    \item Limiting the amount of information in the high $k$-modes results in weaker WDM constraints of $\mWDM$ $<$ 3.6 keV, consistent with previous work by \cite{Villasenor2023}. We further recover similar findings by \cite{Irsic2023} on that the instrumental noise on the data might be underestimated by $\approx$ 30\%.
    \item Modeling of patchy reionization and thermally dependent resolution correction can vary the 2$\sigma$ constraints on $\fWDM$ by $\approx$ 5\% in most cases. Further work on investigating the effect of peculiar velocity on small scales, as pointed out by \cite{Irsic2023}, as well as on a cosmology dependent resolution correction, will help mitigate these differences.  
\end{itemize}
Our results build upon the latest constraints on pure WDM from \cite{Irsic2023}, to further highlight the Lyman-$\alpha$ forest as a unique and sensitive probe of matter clustering on sub-galactic scales. 
Extending this analysis to CWDM cosmologies, we find that the updated constraints have broader implications for any dark matter model that suppresses the amount of clustering on small scales. Our findings suggest that the scale of suppression can be moved to larger scales, but only if the level of the suppression on smaller scales is lower. A logical next step is to construct a general framework that quantifies the allowed suppression amplitude and its scale dependence, enabling comparison across a wide class of dark matter models exhibiting small-scale power suppression (\cite{Archi19, Hooper22, Rogers2021}). A parameterisation of this kind was introduced by \cite{Murgia18} for a large set of both thermal and non-thermal relics. We leave the development of such a generalised framework for future work.

The main interest in constraining this broad class of dark matter models lies in their potential to resolve the $S_{8}$ tension when the Lyman-$\alpha$ forest is combined with other large-scale structure probes (e.g. \cite{Esposito22}). For instance, the tension has been reported to be alleviated when incorporating CMB data in the presence of axion DM (\cite{Rogers25}) or with weak lensing data (\cite{Peters25}). A detailed investigation of the implications for $S_{8}$ within the particular CWDM models allowed in this work is also left for future study.

\bigskip

\acknowledgments 

The authors would like to thank David Chemaly, Tom Hehir, Harry Bevins and Will Handley for helpful conversations.

VI acknowledges support by the Kavli Foundation. MV is supported by the INFN PD51 INDARK grant. Support by ERC Advanced Grant 320596 ‘The Emergence of Structure During the Epoch of Reionization’ is gratefully acknowledged. MGH has been supported by STFC consolidated grant ST/N000927/1 and ST/S000623/1. JSB is supported by STFC consolidated grant ST/X000982/1. For the purpose of open access, the author has applied a Creative Commons Attribution (CC BY) licence to any Author Accepted Manuscript version arising from this submission.

The simulations used in this work were performed using the Joliot Curie supercomputer at the Trés Grand Centre de Calcul (TGCC) and the Cambridge Service for Data Driven Discovery (CSD3), part of which is operated by the University of Cambridge Research Computing on behalf of the STFC DiRAC HPC Facility (https://dirac.ac.uk/). We acknowledge the Partnership for Advanced Computing in Europe (PRACE) for awarding us time on Joliot Curie in the 16th call. The DiRAC component of CSD3 was funded by BEIS capital funding via STFC capital grants ST/P002307/1 and ST/R002452/1 and STFC operations grant ST/R00689X/1. This work also used the DiRAC@Durham facility managed by the Institute for Computational Cosmology on behalf of the STFC DiRAC HPC Facility. The equipment was funded by BEIS capital funding via STFC capital grants ST/P002293/1 and ST/R002371/1, Durham University and STFC operations grant ST/R000832/1. DiRAC is part of the National e-Infrastructure.

\bibliography{references}

\begin{thebibliography}{127}
\bibitem[{foo()}]{footnote}
\bibinfo{note}{The 1D Lyman-$\alpha$ clustering is sensitive only to line-of-sight Fourier modes $k_{\parallel}$. For the sake of simplicity we adopt a shorthand notation of $k$ in velocity units to denote wavenumbers along the line-of-sight.}
\end{thebibliography}

\clearpage

\end{document}